\documentclass[lettersize,journal]{IEEEtran}
\usepackage{amsmath,amsfonts}
\usepackage{algorithm}
\usepackage{array}
\usepackage{textcomp}
\usepackage{stfloats}
\usepackage{url}
\usepackage{verbatim}
\usepackage{graphicx}
\usepackage{cite}
\usepackage{mathptmx}
\usepackage{multirow}
\usepackage{makecell}
\usepackage{graphicx}
\usepackage{threeparttable}
\usepackage{amssymb}
\usepackage{siunitx}
\usepackage{xcolor}
\usepackage{subfigure}
\usepackage{algpseudocode}
\usepackage[hidelinks]{hyperref}
\definecolor{darkred}{RGB}{150,0,0}
\definecolor{darkgreen}{RGB}{0,150,0}
\definecolor{darkblue}{RGB}{0,0,150}
\hypersetup{colorlinks=true, linkcolor=red, citecolor=blue, urlcolor=darkblue}

\begin{document}

\title{Formation and Investigation of Cooperative Platooning at the Early Stage of Connected and Automated Vehicles Deployment}

\author{Zeyu Mu$^{1}$,  Sergei S. Avedisov$^{2}$,  Ahmadreza Moradipari$^{2}$, B. Brian Park$^{3}$
\thanks{This work is supported by the National Science Foundation under Grant No. (CMMI-2009342) and the Toyota Motor North America R\&D.}
\thanks{$^{1}$Z. Mu is with the Link Lab and Systems \& Information Engineering, University of Virginia, Charlottesville, VA 22904 USA; (email: dwe4dt@virginia.edu).}
\thanks{$^{2}$S. S. Avedisov and A. Moradipari are with Toyota Motor North America R\&D - InfoTech Labs, Mountain View, CA, USA; (email: sergei.avedisov@toyota.com; ahmadreza.moradipari@toyota.com).} 
\thanks{$^{3}$B.B. Park is with the Link Lab and Departments of Civil \& Environmental Engineering and Systems \& Information Engineering, University of Virginia, Charlottesville, VA 22904 USA; (email: bp6v@virginia.edu).}}

\markboth{IEEE Transactions on Intelligent Transportation Systems}%
{Shell \MakeLowercase{\textit{et al.}}: A Sample Article Using IEEEtran.cls for IEEE Journals}


\maketitle

\begin{abstract}
Cooperative platooning, enabled by cooperative adaptive cruise control (CACC), is a cornerstone technology for connected automated vehicles (CAVs), offering significant improvements in safety, comfort, and traffic efficiency over traditional adaptive cruise control (ACC). This paper addresses a key challenge in the initial deployment phase of CAVs: the limited benefits of cooperative platooning due to the sparse distribution of CAVs on the road. To overcome this limitation, we propose an innovative control framework that enhances cooperative platooning in mixed traffic environments. Two techniques are utilized: (1) a mixed cooperative platooning strategy that integrates CACC with unconnected vehicles (CACCu), and (2) a strategic lane-change decision model designed to facilitate safe and efficient lane changes for platoon formation. Additionally, a surrounding vehicle identification system is embedded in the framework to enable CAVs to effectively identify and select potential platooning leaders. Simulation studies across various CV market penetration rates (MPRs) show that incorporating CACCu systems significantly improves safety, comfort, and traffic efficiency compared to existing systems with only CACC and ACC systems, even at CV penetration as low as 10\%. The maximized platoon formation increases by up to 24\%, accompanied by an 11\% reduction in acceleration and a 7\% decrease in fuel consumption. Furthermore, the strategic lane-change model enhances CAV performance, achieving notable improvements between 6\% and 60\% CV penetration, without adversely affecting overall traffic flow. 
\end{abstract}


\section{Introduction}

Connected vehicles (CVs) represent a transformative advancement in vehicular technology, enabling the exchange of traffic information both among vehicles and with infrastructure. This capability is key to notable improvements in traffic safety, quality of life, and transportation efficiency \cite{Safety2019Md}. A prominent innovation within this domain is cooperative platooning, specifically cooperative adaptive cruise control (CACC), a control system in connected automated vehicles (CAVs) designed to improve safety, comfort, and energy efficiency \cite{CACC_bef, Aerodynamic2020Kaluva, CAV_r}. CACC also has the potential to increase road capacity and traffic efficiency \cite{Unravelling2018Lin,Modeling2018Hao} without requiring significant infrastructure investments, such as additional lanes or ramp metering. 

In the near future, roadways are anticipated to host both connected vehicles (CVs) and traditional human-driven vehicles (THVs), with the latter being unconnected. A key challenge in this mixed-traffic environment is the risk associated with sudden braking by THVs, which can trigger a phenomenon known as dangerous signal amplification across vehicle strings, compromising traffic flow stability \cite{ACC_resp_T}. Cooperative adaptive cruise control (CACC) offers a promising solution by mitigating this amplification effect, improving upon traditional adaptive cruise control (ACC) \cite{ACC_d}. Extensive research has explored CACC’s impact in mixed traffic, particularly under varying market penetration rates (MPRs) of CAVs, THVs, and combinations of CAVs, connected human-driven vehicles (CHVs), and THVs \cite{van2006The, Steven2012Impacts, Qing2003Simulation}. These studies consistently indicate that higher cooperative platooning levels bring significant benefits to both CAVs and the broader traffic system. 

\begin{figure}[t]
  \centering
  \includegraphics[width=0.4\textwidth]{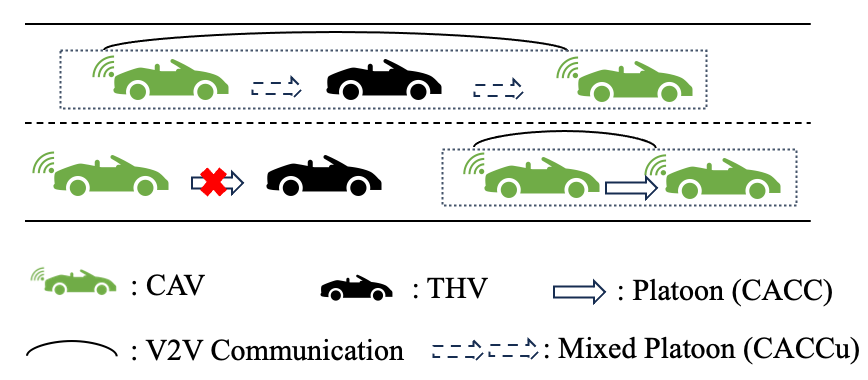}
  \caption{Cooperative platoons in a sparse distribution of connected automated vehicles}\label{fig:sparse}
\end{figure}

However, in the early stages of CAV deployment, the limited and dispersed presence of CAVs reduces the likelihood of one CAV following another, hindering efficient cooperative platoon formation (see Fig. \ref{fig:sparse}). Limited research has fully investigated the potential of the CAV under low MPRs. Existing research on CAV at low MPRs \cite{he_impact_2022, chen2022cooperativeperception, Mart_nez_2023,YAO2024Optimal,systems2025Cheng} has primarily demonstrated performance improvements at moderate MPR levels (typically 20–50\%) or has focused on enhancing ACC-based vehicle behavior through anticipation and multi-vehicle coordination. Although these studies highlight that even limited CAV deployment can yield certain benefits, they often neglect the potential of connectivity-enabled strategies such as CACC. Furthermore, the impact and feasibility of these cooperative strategies at extremely low MPRs (e.g., below 20\%) remain underexplored, despite their critical relevance to near-term deployment scenarios. Recognizing that CACC substantially outperforms ACC in safety, efficiency, and fuel economy even, we propose two key enhancements for low MPR environments: mixed platooning and strategic platoon organization through lane changes.

Mixed cooperative platooning, specifically cooperative adaptive cruise control with an unconnected vehicle (CACCu) \cite{Chen2022Cooperative}, represents an innovative extension of CACC technology. Unlike standard CACC, CACCu does not require the immediately preceding vehicle to be connected; instead, it enables CAVs to cooperate with a connected vehicle further ahead, as illustrated in Fig. \ref{fig:sparse}. This adaptation expands cooperative platooning capabilities in mixed-traffic environments. While prior studies have evaluated CACCu performance at the individual vehicle level \cite{Mu2021Assessment, Mu2022Cooperative}, its broader impact in mixed traffic remains underexplored. Investigating CACCu's dynamics within mixed traffic is crucial for understanding its potential effects on CAVs and overall traffic flow. Integrating CACCu could significantly enhance CAVs' ability to navigate mixed-traffic scenarios, offering a more adaptable approach to cooperative platooning. Thus, a comprehensive examination of CACCu’s role in mixed-traffic simulations is essential to optimize its real-world implementation.

Strategically organizing CAVs within traffic management systems enhances cooperative platooning effectiveness. Several researchers have proposed strategies for organizing CAVs for cooperative platooning. One widely discussed approach to enhance cooperative platooning involves the use of dedicated lanes for CAVs \cite{he_impact_2022,Design2020Solmaz,Shared2022Khaled, Network2024Hua, Chen2024Leveraging,YAO2024Optimal,XIAO2025Lane}. These exclusive lanes can significantly improve platoon formation rates and overall traffic flow by allowing connected vehicles to travel without interference from HDVs. However, the implementation of such lanes during the early stages of CAV deployment remains challenging. At low MPRs, allocating entire lanes for exclusive CAV use can lead to underutilization and inefficient road space usage. To address this limitation, recent studies have proposed dynamic or mixed-use dedicated lanes, where HDVs may be allowed to use the lane but must yield to CAVs \cite{Shared2022Khaled,Chen2024Leveraging,YAO2024Optimal}. While these strategies increase lane utilization, they introduce new challenges such as platoon disruption and the need for advanced coordination mechanisms. Therefore, flexible decisions for these transitional environments that enable CAVs to form stable platoons by dynamically identifying connected leaders and strategically executing lane changes to align with them are needed to offer a more flexible and practical solution during the early deployment phase. Such maneuvers can increase CAV platooning, enhancing road safety, optimizing space, and reducing accidents and energy consumption over time \cite{Lane2006Jorge, Capacity2018Danjue, Freeway2021Hao}. However, frequent lane changes may introduce traffic dynamics complexities, potentially disrupting flow and diminishing platooning efficiency gains. This dual impact presents a significant challenge for traffic management, requiring a balance between maximizing the benefits of cooperative platooning and minimizing potential disruptions.

Establishing cooperative platooning with a connected preceding vehicle or identifying a connected vehicle in an adjacent lane for lane changes relies critically on effective communication between CAVs in real-world scenarios. To form cooperative platoons, CAVs must accurately identify the correct preceding vehicle among nearby connected vehicles, a process that depends on the precision of Global Positioning System (GPS) data received via vehicle-to-vehicle (V2V) communication \cite{V2X}. However, as noted by \cite{Vehicular2008Azzedine}, commercially available GPS devices can exhibit positioning errors of 1-4 meters. Such errors can lead to misidentification, with a CAV incorrectly designating a vehicle in an adjacent lane as its preceding vehicle when driving in parallel, potentially compromising the safety and effectiveness of the cooperative platooning system.  Existing traffic simulation studies \cite{Efficient2017Clark, Modelling2019Jian, A2021Md, Connected2021Rasool, Li2022Simulation, Wang2022Wang} often overlook the time and accuracy involved in identifying the correct leader for seamless engagement in CACC or other cooperative maneuvers. Without adequately accounting for leader identification efficiency, simulation results may not accurately reflect real-world scenarios, limiting their reliability for practical estimation.

Given the research gaps outlined, this paper introduces a comprehensive framework to explore and maximize the potential benefits of cooperative platooning in the early stages of CAV deployment within mixed traffic. This framework is built around several key systems. First, a surrounding vehicle identification system is implemented to accurately identify connected candidate vehicles nearby for cooperative platooning. Once a potential leader is identified, various control planners, including ACC, CACC, and CACCu can be employed for platooning. Second, a strategic lane change decision model for cooperative platoons is developed to maximize platoon cohesion. To assess the benefits of the proposed framework on automated vehicles and overall traffic flow, we conduct simulation experiments across various MPRs of automated vehicles, testing four types for comparison: 1) automated vehicles (AVs) with only ACC functionality; 2) connected automated vehicles (CAVs) with ACC and CACC capabilities; 3) connected automated vehicles with unconnected vehicles (CAVu) that support ACC, CACC, and CACCu; and 4) connected automated vehicles with unconnected vehicles and lane change capabilities (CAVu-LC), utilizing ACC, CACC, CACCu, and strategic lane changes.

The main contributions of this paper are listed as follows:
\begin{itemize}
\item Developed a comprehensive CAV control framework to enhance cooperative platooning in traffic scenarios with limited connectivity. The framework integrates a practical vehicle identification procedure, dynamically selects appropriate control strategies (ACC, CACC, and CACCu), and identifies potential leaders to facilitate strategic lane changes, enabling effective and adaptive platoon formation.
\item Performed high-fidelity microsimulations to demonstrate the substantial benefits of the proposed framework for automated vehicles and overall traffic performance, even with low connected vehicle penetration rates. The analysis emphasizes improvements in driving comfort, energy efficiency, and traffic flow efficiency.
\item Investigated the role and impact of vehicle identification systems in cooperative platoon formation within mixed traffic environments, highlighting their critical influence on CAV performance and overall traffic dynamics. This study provides valuable insights for evaluating and optimizing automated vehicle systems in real-world applications.
\end{itemize}

The remainder of the paper is organized as follows: Section \ref{sec:framework} Control Framework of Automated Vehicles provides an overview of the designed CAV control framework aimed at maximizing cooperative platooning. Section \ref{sec:microsimulation} Microsimulation for Mixed Traffic describes the high-fidelity microsimulation for mixed traffic flow, including a human car-following model with human factors, vehicle dynamics modeling, and relevant data inputs. Section \ref{sec:design} Experimental Design and Configuration outlines the experiment designs, traffic flow data, assumptions, simulation parameters, and performance metrics. Section \ref{sec:results} Simulation Results and Analysis presents the performance of the surrounding vehicle identification system along with the simulation results and analysis. Finally, the Section \ref{sec:conclusion} Conclusions and Future Works offers concluding remarks and directions for future research.

\section{Control Framework of Automated Vehicles} \label{sec:framework}

In mixed-traffic scenarios, CAVs must accurately identify both preceding and surrounding vehicles to implement cooperative strategies effectively. Specifically, a CAV can engage in cooperative platooning when following a CV; it can also initiate mixed cooperative platooning by coordinating with a remote connected vehicle when following a THV. Additionally, if the immediate preceding vehicle is unconnected, the CAV may execute a lane change to follow a connected vehicle in an adjacent lane. To facilitate such flexible platooning in real-world mixed traffic, we developed a control framework integrating a Surrounding Vehicle Identification System (SVIS) \cite{SVIS}, a strategic lane-change model for cooperative platooning, and multiple control planners. This control framework is designed to optimize CAV capabilities for cooperative platooning, promoting flexible and efficient interactions in mixed-traffic environments.

\begin{figure}[t]
  \centering
  \includegraphics[width=0.4\textwidth]{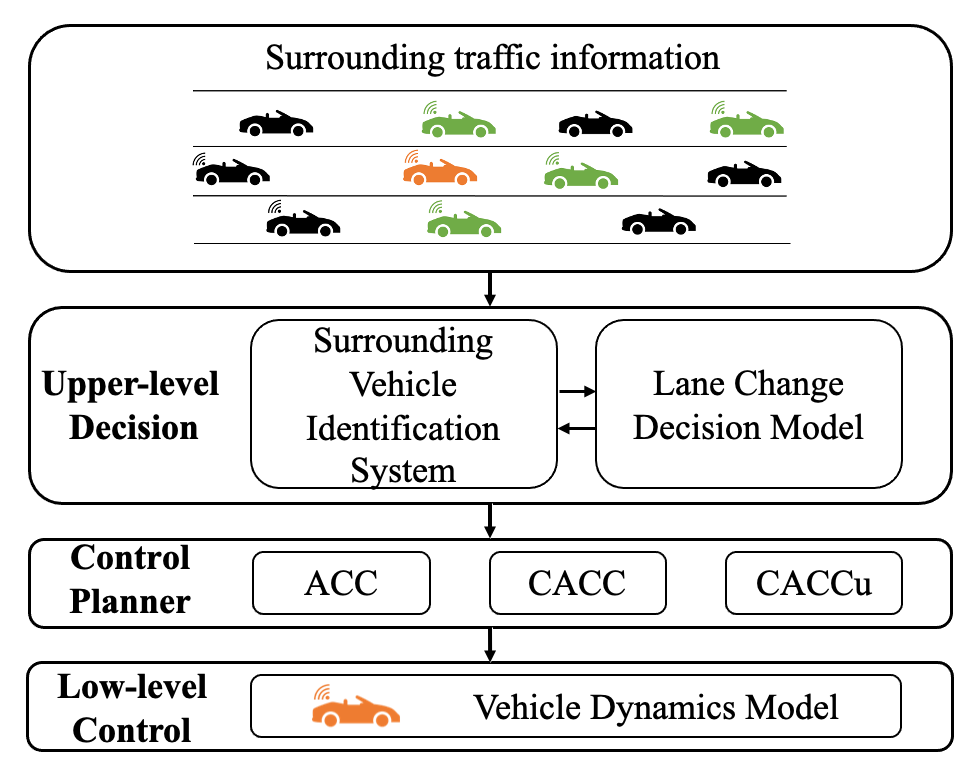}
  \caption{Control Framework of Connected Automated Vehicles}\label{fig:Framework}
\end{figure}

The CAV control framework consists of three main components: the upper-level decision module, the control planner, and the low-level controller, as outlined in Fig. \ref{fig:Framework}. The upper-level decision module includes a SVIS for identifying nearby connected vehicles suitable for cooperative platooning, along with a lane change decision model that uses SVIS output to assess the feasibility of changing lanes to follow a connected vehicle. In this study, we assume that identifying a second preceding vehicle requires approximately twice the time needed for identifying the immediate preceding vehicle. This assumption is based on the fact that detecting a second preceding vehicle involves additional sensing and data fusion steps, as the signal must pass through or around the first vehicle. Consequently, the accumulated uncertainty from radar and GPS measurements, such as range and position errors, can roughly double due to the increased sensing distance and reliance on multiple sensor inputs. Based on the connectivity status of the first and second preceding vehicles, the control planner then selects the appropriate control strategy (ACC, CACC, or CACCu). Finally, the low-level controller executes the control commands generated by the control planner. The details of the control framework are explained in this section. Table \ref{tab:notations} summarizes all key symbols and parameters used throughout the study.

\begin{table}[!t]
\centering
\caption{List of Notations Used in This Study}
\scriptsize
\begin{tabular}{lll}
\hline
\textbf{Symbol} & \textbf{Description} & \textbf{Unit / Type} \\
\hline
$x_i, \dot{x}_i, \ddot{x}_i$ & Position, speed, acceleration of vehicle $i$ & m, m/s, m/s$^2$ \\
$h_i = x_{i-1} - x_i$ & Distance to preceding vehicle & m \\
$e(t)$ & Spacing error & m \\
$u(t)$ & Control input (acceleration) & m/s$^2$ \\
$v_i$ & Speed of vehicle $i$ & m/s \\
$\Delta d_i, \Delta v_i$ & Gap and relative speed & m, m/s \\
$\Delta d_{i,d}$ & Desired gap & m \\
$TTC(t)$ & Time-to-collision & s \\
$\tilde{a}, \tilde{a}_n, \tilde{a}_o$ & Accelerations after lane change & m/s$^2$ \\
$C, P_i$ & Set of CAVs; preceding vehicles & set \\
$con_{ij}$ & Connectivity status (0/1) & binary \\
$S_{ij}$ & Cooperative platoon size & count \\
$LC$ & Lane-change decision (1/0) & binary \\
$T_h$ & Desired time headway & s \\
$k_p, k_d$ & PD gains & -- \\
$f(\ddot{x}_{i-1})$ & Feed-forward (CACC) & -- \\
$f'(\ddot{x}_{i-1})$ & Feed-forward (CACCu) & -- \\
$\alpha, \beta, T_{\text{ovm}}$ & OVM gains, headway & --, s \\
$\tau, t_d$ & Powertrain constant, delay & s, s \\
$b_{\text{safe}}$ & Safe deceleration (MOBIL) & m/s$^2$ \\
$p$ & Politeness factor & -- \\
$\Delta a_{\text{th}}$ & Incentive threshold & m/s$^2$ \\
$T, v_{i,d}, s_0, a, b$ & Headway, speed, gap, accel., decel. & s, m/s, m, m/s$^2$, m/s$^2$ \\
$b_e$ & Emergency deceleration & m/s$^2$ \\
$k$ & Reaction delay & s \\
$v_s, \sigma_r, \tau$ & Error/persistence parameters & --, --, s \\
$\delta_x, \delta_y, \delta_v$ & GPS/radar errors & m, m, m/s \\
$\alpha_1, \alpha_2$ & Region-test parameters & -- \\
$n, k$ & Iteration indices & count \\
$E_r, U_r$ & Misident. and unusability rates & \% \\
$T$ & ID time for second vehicle & s \\
\hline
\end{tabular}
\label{tab:notations}
\end{table}

\subsection{Surrounding Vehicle Identification System}

Traffic environments are highly dynamic, requiring CAVs to continuously identify which vehicles are connected and suitable for cooperative maneuvers such as platooning or lane changes. Given common GPS distance errors ranging from 1 to 4 meters \cite{Vehicular2008Azzedine}, it is essential for CAVs to accurately and promptly identify the correct preceding vehicle or lane-change target to initiate cooperation effectively.  Unlike most mixed traffic simulations, which often overlook the timing and accuracy of vehicle identification for CAV cooperation, our framework incorporates a Surrounding Vehicle Identification System (SVIS) \cite{SVIS} to account for real-world factors such as GPS inaccuracies and radar sensor limitations. By modeling the dynamic and imperfect nature of the identification process, SVIS enhances the realism and reliability of the identification procedures. In addition, the SVIS prototype has demonstrated promising performance in preliminary real-world testing \cite{mu_prototype_2024}.

The core function of the SVIS system is to iteratively match the relative position and speed data of the preceding vehicle, obtained from the sensor system, with GPS information from wireless messages sent by connected surrounding vehicles, as illustrated in Fig. \ref{fig:SVIS}. This iterative matching process is crucial for accurately identifying the appropriate leader among nearby connected vehicles for cooperative platooning in real-world scenarios. To prioritize both computational efficiency and robust performance and to make it well-suited for practical deployment, three main factors are considered:
\begin{figure}[t]
  \centering
  \includegraphics[width=0.5\textwidth]{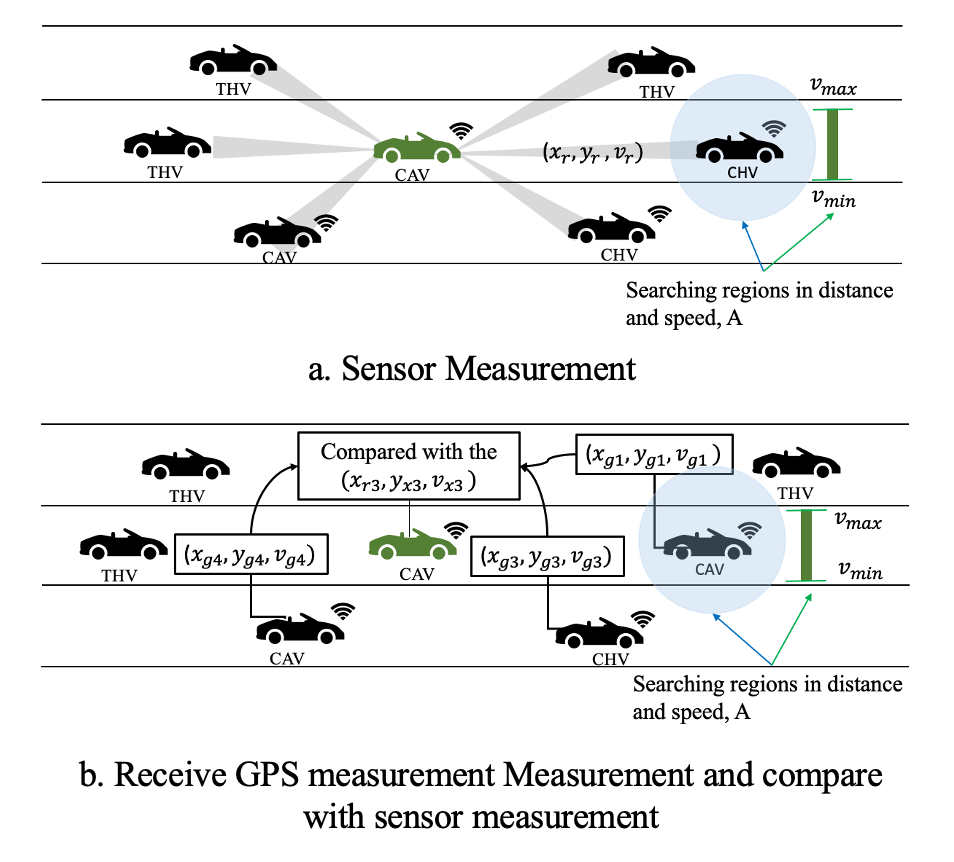}
  \caption{Matching GPS information from nearby connected vehicles with radar information of the actual target surrounding vehicle}\label{fig:SVIS}
\end{figure}

\begin{itemize}
    \item Misidentification error ($Er$): The percentage of nearby connected vehicles incorrectly identified as target connected vehicles out of all identified vehicles.
    \item Unusability rate ($Ur$): The percentage of connected vehicles incorrectly classified as unconnected out of all identified vehicles.
    \item Time consumption ($T$): The time required to complete the identification procedure.
\end{itemize}

To address the three key factors mentioned above, the identification matching procedure for target surrounding vehicles is optimized based on the error distribution of GPS and radar measurements, as well as acceptable accuracy and time efficiency constraints. The position and speed regions for matching serve as thresholds to evaluate whether the difference between GPS data from nearby connected vehicles and sensor data from target surrounding vehicles falls within acceptable limits. The defined position and speed regions, illustrated in Figs. \ref{fig:SVIS} (a) and \ref{fig:SVIS} (b), respectively, are governed by Eqs. (\ref{eq:SVIS_thrh_d}) and (\ref{eq:SVIS_thrh_v}).

\begin{align}
(\frac{e_{gx}-e_{rx}}{\delta_x})^2 + (\frac{e_{gy}-e_{ry}}{\delta_y})^2 \, < \, \text{inv}\chi^2(2,\alpha_1) \label{eq:SVIS_thrh_d} \\
(\frac{e_{g}-e_{r}}{\delta_v})^2 \, < \, \text{inv}\chi^2(1,\alpha_2) \label{eq:SVIS_thrh_v}
\end{align} 
where $e_{gx}$ and $e_{gy}$ represent the errors in GPS longitudinal and lateral distance measurements, defined as the difference between the GPS-measured distance and the ground truth, while $e_r$ denotes the error in radar speed measurement. In this study, "speed" primarily refers to the longitudinal velocity of the vehicle, with lateral velocities, even during lane changes, considered negligible in highway driving contexts. According to GPS error modeling studies \cite{GPS_error}, these measurement errors follow a normal distribution. Additionally, $\delta_x$ and $\delta_y$ are the standard deviations of independent normal variables representing GPS errors in the longitudinal and lateral directions, respectively, and $\delta_v$ is the standard deviation of the radar speed error. The parameters  $\alpha_1$ and $\alpha_2$ are the probabilities of the target surrounding vehicle being outside the distance and speed regions, respectively, while $\text{inv}\chi^2$ denotes the inverse Chi-square distribution. Note that the GPS speed error is typically much smaller than, and is insignificant compared to, the radar speed error, according to \cite{GPS_error1,GPS_error2}, an assumption reflected in Eq. (\ref{eq:SVIS_thrh_v}).

With the information from radars and GPS, the identification procedure is conducted with an inner loop to identify the connected preceding candidates for $n$ times  and an outer loop to identify the unconnected vehicle by repeating the inner loop for $k$ times. The connected preceding vehicle is considered identified if the speed and distance measurements from the vehicle's radars match the respective GPS measurements within the speed and position region for $n$ consecutive steps, and the unconnected preceding vehicle is considered identified if the matching does not satisfy the speed and position region during $k$ times. The parameters $n$, $k$ and the position and speed regions are optimized based on Eqs. (\ref{eq:SVIS_thrh_d}) and (\ref{eq:SVIS_thrh_v}) and requirements refer to study \cite{SVIS}.

\subsection{Lane Change Decision Model for Cooperative Platooning}

\begin{figure}[t]
  \centering
\includegraphics[width=0.3\textwidth]{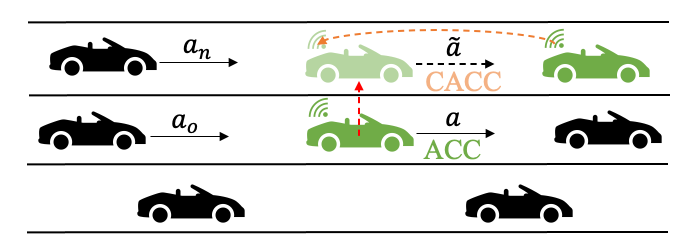}
  \caption{Lane change for cooperative platoon}\label{fig:MOBIL}
\end{figure}

In mixed traffic scenarios with a low market penetration rate of connected vehicles, the likelihood of encountering a CV is limited. In such cases, executing lane changes to follow a connected vehicle in an adjacent lane offers a promising approach to achieving long-term benefits in safety, comfort, and energy efficiency. However, enabling safe lane changes for cooperative platoons requires a lane change decision model that not only accounts for the advantages of cooperative platooning but also prioritizes safety during the lane change and maintains traffic efficiency afterward. To address this, we introduce a lane change decision model based on the Minimizing Overall Braking Induced by Lane Changes (MOBIL) principle, specifically tailored for cooperative platooning, as illustrated in Fig. \ref{fig:MOBIL}.

MOBIL \cite{MOBIL} guides lane change decisions by evaluating the incentive and associated risk. This model defines the safety of a lane change by assessing the change in longitudinal acceleration experienced by the follower in the target lane. The primary incentive for executing a lane change is to enhance overall traffic conditions, either by allowing the vehicle to move at a faster speed or by avoiding slower-moving leaders. The mathematical formulation of MOBIL is detailed in Eq. (\ref{eq:MOBIL_safe}) for safety and Eq. (\ref{eq:MOBIL_incentive}) for incentives.

\begin{align}
\tilde{a}_n(t) > b_{\text{safe}} \label{eq:MOBIL_safe} \\
\tilde{a}(t) - a(t) + p \, (\tilde{a}_n(t) - a_n(t) + \tilde{a}_o(t) - a_o(t)) > \vartriangle a_{\text{th}} \label{eq:MOBIL_incentive}
\end{align} 

where $a$ is the acceleration of the ego vehicle in the current lane and $\tilde{a}$ is its acceleration after the prospective lane change; $a_n$ and $\tilde{a}_n$ are the acceleration of the new follower of ego vehicle before and after the ego vehicle's prospective lane, respectively; $a_o$ and $\tilde{a}_o$ are the acceleration of the old follower of ego vehicle before and after the ego vehicle's prospective lane, respectively; It is noted that the acceleration after a prospective lane change is estimated by the IDM \cite{IDM}. The parameter $b_{\text{safe}}$ is the maximum acceptable deceleration, and $p$ is a politeness parameter, which balances the deceleration of other vehicles with the gain in the subject vehicle's own acceleration. The threshold $\vartriangle a_{\text{th}}$ is designed to prevent lane changes if the overall benefit is only marginal compared to maintaining the current lane.

\begin{algorithm} [t]
\caption{Lane Change Decision Model for Cooperative Platooning}\label{alg:LC}
\begin{algorithmic}[1]
\State Initialize $LC_{ij} \gets 0$, $S_{ij} \gets 0$  for $i \in C$ and $j \in \{0, 1, 2\}$,  $\text{con}_{ij} \gets$ SVIS
\For{$i$ in $C$}
\For{$j = 0$ to $2$} 
    \For{$k \in P_i$} 
        \State $\text{S}_{ij} \gets \text{S}_{ij} + 1$ If $\text{con}_{k0} == 1$ else break
    \EndFor
\EndFor
\For{$j = 1$ to $2$} 
    \If{$\text{con}_{i0} = 0$ and $\text{con}_{ij} > 0$} 
        \State $B_{ij} \gets f(S_{ij}, \text{Eq.} \ref{eq:MOBIL_safe}, \text{Eq.} \ref{eq:MOBIL_incentive})$
    \Else
        \State $B_{ij} \gets None$
    \EndIf
\EndFor
\State $B_i = \max(B_{i1}, B_{i2})$ if  $B_{ij} \neq None$
\For{$j = 1$ to $2$}
    \State $LC(j) \gets 
        \begin{cases} 
        1, & \text{if } B_{ij} \neq None \text{ and }  B_{ij} = B_i \\
        0, & \text{otherwise}
        \end{cases}$
\EndFor
\EndFor
\end{algorithmic}
\end{algorithm}

Building upon the MOBIL framework, which assesses safety and speed benefits, we have integrated additional criteria into our lane change decision model: the connectivity of prospective preceding vehicles and the size of the cooperative platoon, which are gained from the SVIS. This enhancement aims to prevent scenarios in which a CAV executing CACC selects an unconnected vehicle on nearby as its leader for a higher speed, thereby reverting to a standard ACC system. Additionally, it facilitates the integration of a CAV into a larger cooperative platoon, maximizing the potential benefits of cooperative driving. The modified lane change decision algorithm for cooperative platooning, which incorporates these considerations, is outlined in Algorithm \ref{alg:LC}. In this algorithm, set $C$ contains all the connected automated vehicles and $P_i$ contains all the preceding vehicles of the vehicle $i\in C$. The connectivity status of the surrounding vehicle of the vehicle $i$ is represented as $\text{con}_{ij}$, with 0 for the unconnected vehicle and 1 for the connected vehicle, and $j = 0, 1, 2$ with respect to the current, left, or right lanes. The set $S_{ij}$ contains the surrounding cooperative platoon size of the vehicle $i$, which is the number of connected vehicles in the front, left, and right lanes. The function $f(\cdot)$ is the lane change criteria that are incorporated through the number of cooperative platoon sizes, Eqs. (\ref{eq:MOBIL_safe}) and (\ref{eq:MOBIL_incentive}). Set $LC$ denotes the lane change decision, with 1 indicating a lane change and 0 indicating retention of the current lane.

\subsection{Control Planner}

\begin{figure}[t]
  \centering
  \includegraphics[width=0.4\textwidth]{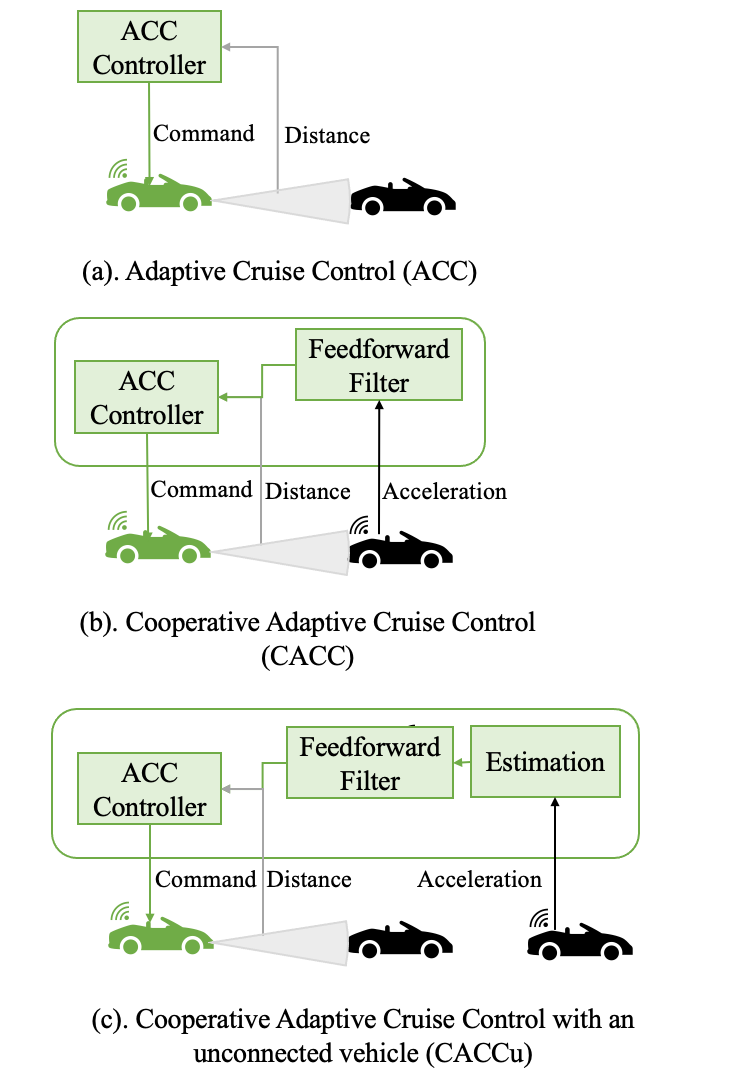}
  \caption{Three control planners: ACC system, CACC system and CACCu system}\label{fig:control_mode}
\end{figure}

\begin{figure}[t]
  \centering
  \includegraphics[width=0.5\textwidth]{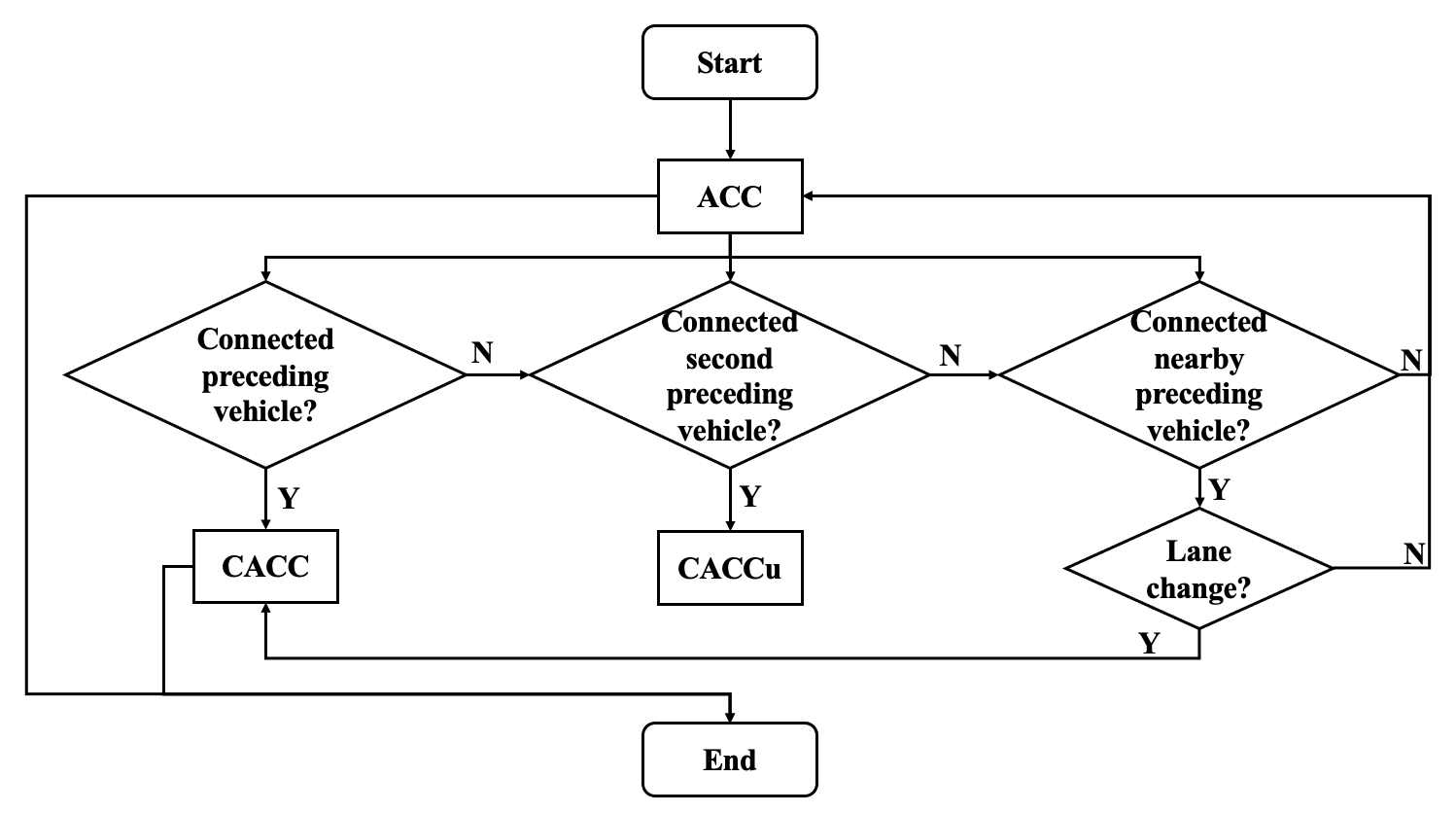}
  \caption{Flowchart of choosing control planner }\label{fig:CAV_Framework}
\end{figure}

This section outlines three types of control planners (i.e., ACC, CACC and CACCu systems), as shown in Fig. \ref{fig:control_mode}. The choice of a particular control planner depends on the connectivity status of both the first and second preceding vehicles in the current lane, as illustrated in Fig. \ref{fig:CAV_Framework}. If the ego vehicle lacks connectivity capabilities or the preceding vehicle is unconnected, the ACC control planner \cite{ACC_d} is utilized, operating without cooperative control. When the ego vehicle has connectivity capabilities and the preceding vehicle is connected, the CACC control planner \cite{ACC_d} is chosen to enable cooperative platooning. In cases where the preceding vehicle is unconnected but the second preceding vehicle is connected, the CACCu control planner \cite{CACCu} is utilized to achieve mixed cooperative platooning. The algorithms governing these three control planners, illustrated in Fig. \ref{fig:CAV_Framework}, are detailed in this section.

The ACC system \cite{ACC-dv} utilizes the sensor-measured distance to keep a safe distance from the preceding vehicle, as shown in Fig. \ref{fig:control_mode}. It employs a proportional-derivative (PD) controller to generate the acceleration command $u(t)$ and stabilizes the spacing error $e(t)$ by utilizing the relative distance ($h_i = x_{i-1} - x_{i}$) from the preceding vehicle and the desired time headway ($T_{h}$). Specifically, the position of the ego vehicle $i$ is defined as $x_i$, and the speed and acceleration are defined as $\dot{x}_i$ and $\ddot{x}_i$, respectively. The acceleration command of the vehicle $i$ is given by Eqs. (\ref{eq:ACC-e}) and (\ref{eq:ACC-u}).
\begin{align}
e(t) = h_{i}(t) - T_h \dot{x}_i(t) \label{eq:ACC-e}\\ 
u(t) = k_p e(t) + k_d \dot{e}(t) \label{eq:ACC-u} 
\end{align}
where $k_p$ and $k_d$ are proportional and derivative gains, respectively. 




Despite the ACC system being equipped in commercially-used automated vehicles, safety remains a concern when following vehicles in a short time headway \cite{Impacts2021Mingfeng}. When an ACC vehicle follows an unconnected preceding vehicle in a short desired time headway, the oscillation caused by a disturbance, such as the harsh deceleration of the lead vehicle, can be amplified along the cooperative platoon. This can lead to congestion and collisions in extreme cases. To address the limitations of the ACC system, the CACC system \cite{ACC_d} is developed to enhance its performance by utilizing additional acceleration information via V2V communication. This requires the first preceding vehicle to be connected, as shown in Fig. \ref{fig:control_mode}. The additional acceleration information allows the CACC vehicle to mitigate disturbances coming from the preceding vehicle with a designed feed-forward filter. The acceleration of the connected preceding vehicle serves as a feed-forward signal through the feed-forward filter $f(\ddot{x}_{i-1})$. The acceleration command of the CACC system is defined by Eq. (\ref{eq:CACCd-U}):
\begin{equation}
u(t) = k_p e(t) + k_d \dot{e}(t) + f(\ddot{x}_{i-1}(t)) \label{eq:CACCd-U}
\end{equation} 
where $e(t)$, $k_p$, $k_d$ are the same as the ACC system, referring to Eqs. (\ref{eq:ACC-e}) and  (\ref{eq:ACC-u}).  
The design of the feed-forward filter for CACC is based on a zero-error condition \cite{ACC_d}.

The foundational requirement for implementing CACC is the presence of a connected preceding vehicle. However, achieving a fully connected environment is projected to take decades. In the interim, particularly in mixed traffic conditions with a low MPR of CVs, CAVs are likely to frequently encounter unconnected preceding vehicles. As a result, the majority of CAVs will need to operate using standard ACC systems.

To overcome this challenge, a CACCu system has been developed \cite{Chen2022Cooperative} and also evaluated in field tests \cite{LEE2021Design}. This system cleverly leverages the second connected vehicle when the first preceding vehicle is unconnected, as shown in Fig. \ref{fig:control_mode}. Like CACC, the CACCu system incorporates a feed-forward filter to stabilize the system. Unlike CACC, CACCu utilizes the acceleration of the second preceding vehicle instead of the first. The acceleration of the first preceding vehicle is estimated using a linearized optimal velocity model (OVM) \cite{Yūki1999Optimal}. The acceleration command is calculated similar to CACC, as defined in Eq. (\ref{eq:CACCd-U}).
\begin{align}
u(t) = k_p e(t) + k_d \dot{e}(t) + f'(\ddot{x}_{i-1} (t)) \label{eq:CACCu-U}
\end{align} 
The feed-forward filter $f_i'(\cdot)$ is calculated based on the zero-error condition and the prior knowledge of human drivers \cite{Chen2022Cooperative}.

\section{Microsimulation for Mixed Traffic} \label{sec:microsimulation}

This section identifies key areas essential for developing mixed traffic simulation models aimed at evaluating automated vehicles. It contains critical elements such as human factor integration in driver driving modeling and vehicle dynamics modeling to ensure a realistic and accurate representation of vehicle responses under diverse driving conditions and control inputs as well as the traffic flow data designed for evaluation of long travel benefits of cooperative platooning. The section summarizes the fundamental conditions and constraints that should inform the development of simulation models that accurately reflect the interactions between human-driven and automated vehicles in mixed traffic scenarios.

\subsection{Modeling Human-Driven Vehicle Behavior} \label{sec:human}
 
Human drivers are pivotal elements in mixed-traffic simulations. The standard Intelligent Driver Model (IDM) \cite{IDM} has been a widely-used car-following model for human drivers, as documented in several studies \cite{Rolling2014Meng,Modeling2014Vicente,Assessing2021Anshuman,Impacts2021Mingfeng,Stability2023Jie}. However, research has highlighted the limitations of the standard IDM in capturing the nuanced behaviors of human drivers accurately \cite{Does2023Zeyu, moradipari2022predicting}. To address these shortcomings, we enhanced the IDM by incorporating human factors, such as reaction delay and estimation errors, aiming to reflect varying levels of aggressiveness in human driving behaviors. These additions render the human model a more realistic representation of actual driving scenarios \cite{moradipari2022predicting,Does2023Zeyu}.
 
\subsubsection{Intelligent Driver Behavior Modeling}
The Intelligent Driver Model (IDM) is a widely used car-following model that describes the acceleration of a vehicle as a function of its speed, the spacing from the preceding vehicle, and the relative speed between the two vehicles. This model effectively captures the car-following behavior by integrating various parameters to simulate realistic driving dynamics.

The mathematical formulation of IDM's predicted acceleration for the $i^{th}$ driver is represented by the following equations:
\begin{align}
        a_{i}(t) = a \left[1 - (\frac{v_{i}(t)}{v_{i,d}(t)})^4 - (\frac{\vartriangle d_{i,d}(v_{i},\vartriangle v_{i}, t)}{\vartriangle d_{i}(t)})^2\right] \label{eq:IDM1} \\
        \vartriangle d_{i,d}(v_{i},\vartriangle v_{i}, t) = s_0 + max\left( v_{i}(t)*T + \frac{v_{i}(t) \cdot \vartriangle v_{i}(t)}{2\sqrt{a b}},\,0\right)
\label{eq:IDM2}
\end{align}
where parameters $T, v_{i,d}, s_0, a$, and $b$ are the safe time headway, desired speed, minimum distance, maximum acceleration, and maximum deceleration, respectively. At time $t$, $a_i$ is the predicted IDM acceleration; $\vartriangle d_{i,d}$ is the desired gap; $\vartriangle d_{i}$ is the space gap; $v_i$ is the speed of vehicle $i$; $\vartriangle v_{i}$ is the relative speed.

\subsubsection{Modeling Human Reaction Delay and Safety Considerations} 
Human drivers typically exhibit reaction delays when following other vehicles. To simulate human-like behavior in traffic simulations involving mixed traffic scenarios, these reaction delays are incorporated. However, excessively long delays can increase safety risks, potentially leading to collisions and resulting in unrealistically aggressive driving behaviors. A commonly used metric to evaluate safety and avoid collisions is Time-to-Collision (TTC), which estimates the time required for two consecutive vehicles in the same lane to collide if they continue at their current speeds. The metric TTC is defined in Eq. (\ref{eq:TTC}) as follows:

\begin{equation}
TTC(t) =
\begin{cases}
\frac{x_{i-1}(t) - x_i(t)}{v_i(t) - v_{i-1}(t)} & \text{if $v_{i-1}(t) < v_i(t)$}\\
\infty & \text{if $v_{i}(t) \leq v_{i-1}(t)$}
\label{eq:TTC}
\end{cases}
\end{equation}

To address safety issues associated with large human reaction delays, the IDM incorporates normal and emergency driving modes based on a predefined TTC threshold, as examined in the study by \cite{Avedisov2022Impacts}. In normal car-following mode, human drivers experience a consistent delay of $k$ seconds. When the TTC value falls below the predefined threshold, indicating an imminent risk of collision, the system switches to an emergency braking mode. In this mode, the delay is eliminated, and an emergency deceleration ($b_e$) is applied immediately. The switch between these two modes is dependent on the driver’s safety awareness and the real-time assessment of collision risks, allowing drivers to react swiftly to evolving traffic conditions.

\subsubsection{Human Estimation Error Modeling}
In this study, we consider both traditional human-driven vehicles, which lack communication capabilities with other vehicles, and connected human-driven vehicles that can exchange state information (e.g., speed and position) with other connected vehicles. To account for human error in following behavior, estimation errors are incorporated into the standard IDM model for both traditional human-driven vehicles and connected human-driven vehicles when following an unconnected vehicle. This approach reflects typical perceptual inaccuracies in gauging the motion of preceding vehicles. In contrast, connected human-driven vehicles can access accurate information when following a connected vehicle, mitigating such perceptual errors. 

The estimation error model for human-driven vehicles is adopted from \cite{Delays2006Martin}, which utilizes a Wiener process to represent common human underestimations of gap and speed differences. The estimated gap ($\vartriangle \tilde{d}{i}$) and estimated speed differences ($\vartriangle \tilde{v}{i}$) are defined as Eqs. (\ref{eq:error1}), (\ref{eq:error2}), and (\ref{eq:error3}):

\begin{equation}
\vartriangle \tilde{d}{i}(t) = \vartriangle d{i}(t) e^{V_s w(t)}; \label{eq:error1}
\end{equation}
\begin{equation}
  \vartriangle \tilde{v}{i}(t) = \vartriangle v{i}(t) s \sigma_r w(t); \label{eq:error2}
\end{equation}
\begin{equation}
 w(t+\vartriangle t) = e^{\frac{\vartriangle t}{\tau}} w(t)+\sqrt{\frac{2\vartriangle t}{\tau}}\eta_t \label{eq:error3}
\end{equation}

Here, $\vartriangle d_{i}(t)$ and $\vartriangle v_{i}(t)$ represent the actual gap and speed difference, respectively. In the Wiener process, $w(t + \Delta t) - w(t)$ is normally distributed with a mean of 0 and a variance of $\Delta t$. For this study, $\Delta t = 1$, making $w(t + \Delta t) - w(t) \sim \mathcal{N}(0, 1)$. The coefficient of variation $v_s$ describes the relative standard deviation between estimation and truth; $\sigma_r$ is the constant standard deviation of the relative approach rate; $\eta_t$ represents independent realizations of a Gaussian-distributed variable with zero mean and unit variance; and $\tau$ is the persistence or correlation time.

\subsection{Vehicle Dynamics Modeling}
The vehicle dynamics model is a mathematical representation of a vehicle's physical behavior, including its response to control inputs such as acceleration, braking, and steering under various road and environmental conditions. This model is essential in automated vehicle simulation as it captures realistic vehicle movement and interactions with external forces, which are crucial for accurately testing and validating control algorithms and assessing AV performance.

In this study, a simplified first-order vehicle dynamics model is used to approximate the longitudinal dynamics of the vehicle, as detailed in \cite{ACC_d}. As this study primarily focuses on highway scenarios, lateral vehicle dynamics are considered negligible. The vehicle dynamics model is defined by Eq. (\ref{eq:veh_dyn}):
\begin{equation}
u(t) = \tau \dot{v}_i(t-t_d) + v_i (t-t_d)
\label{eq:veh_dyn}
\end{equation}
In this expression, $\tau$ represents the power train time constant and $t_d$ is the system delay. 

\subsection{Extended NGSIM-Based Simulation for Long-Distance Travel}

Real-world traffic data from the Next Generation Simulation (NGSIM) US-101 highway dataset \cite{NGSIM} is utilized to model traffic flow. The NGSIM dataset includes 45 minutes of high-resolution driving data, recorded at 10 Hz, and covers a segment of approximately 640 meters in Los Angeles. For our simulation, we selected the first 15 minutes of the dataset, which encompasses about 2000 vehicles and represents non-congested traffic conditions.

\begin{figure}[t]
\centering
    \subfigure[Choosing trajectories from data]{\includegraphics[width=1\linewidth]{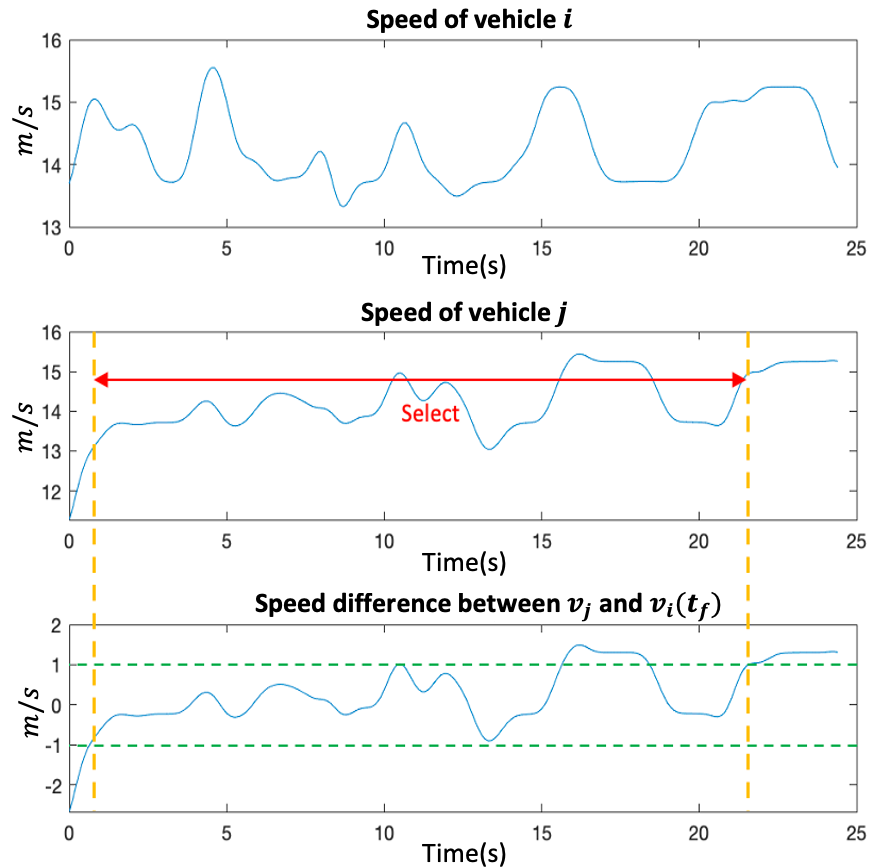}}
    
    \subfigure[Acceleration and speed transition]{\includegraphics[width=1\linewidth]{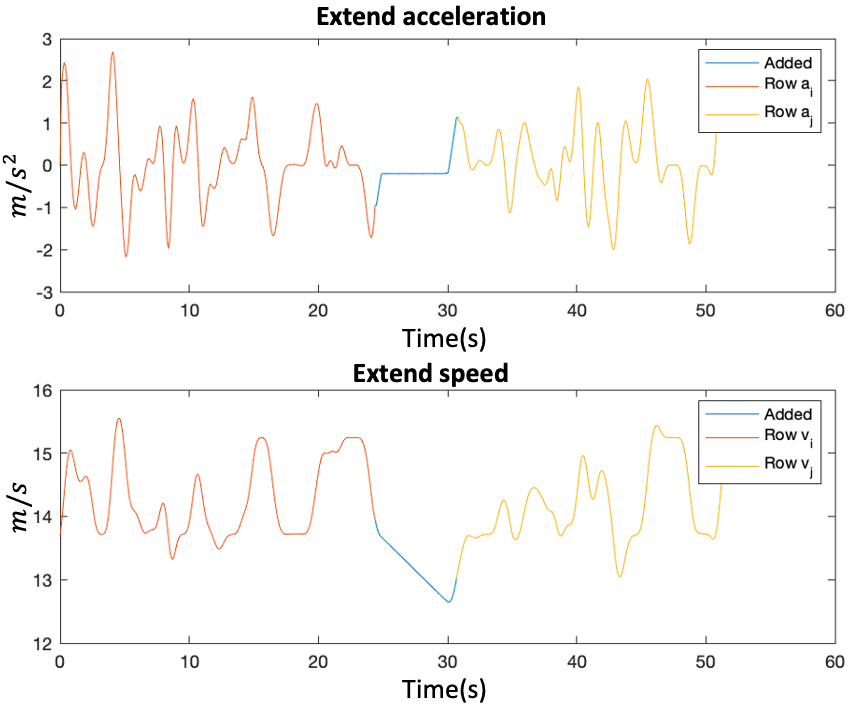}}
    \caption{An example of extended human trajectory}
    \label{fig:extend_NGSIM}
\end{figure}

Given the limited coverage of approximately 640 meters in the NGSIM data, we extended the lead vehicle trajectories to evaluate and demonstrate the long-term benefits of cooperative platooning through the use of CACCu for mixed traffic platooning and lane changes for CACC in scenarios with low MPR of CAVs. Specifically, we extended the lead vehicle trajectory on each lane by combining multiple trajectories from vehicles in the same lane without lane changes, as recorded in the NGSIM data. To seamlessly connect different lead vehicle trajectories within each lane, we followed these steps and showed an example of extended trajectories shown in Fig. \ref{fig:extend_NGSIM}.

\begin{enumerate}
\item Exclude vehicles that make lane changes. For the remaining vehicles, sort the speed trajectories based on the time they enter the road segment.
\item Choose the trajectories of the vehicle $i$ and $j$. Subtract the final speed of the vehicle $i$ from the speed trajectory of the vehicle $j$ to determine the speed difference $\triangle v_i(t)$.
\item Select the trajectory between the first and last instance when $\triangle v_i(t)$ is within 1 m/s to extend the trajectory, as shown in Fig. \ref{fig:extend_NGSIM}(a).
\item To connect the final speed of the vehicle $i$ with the initial speed of vehicle $j$, apply a fixed acceleration/deceleration of $0.2~\mathrm{m/s^2}$ and a rate of $0.2 ~\mathrm{m/s^3}$.
\item If the final speed of the vehicle $j$ is higher than the initial speed of the vehicle $i$, adjust the final acceleration/deceleration of the vehicle $j$ and the initial speed of the vehicle $i$ to $0.2 ~\mathrm{m/s^2}$ to achieve a smooth transition, as shown in Fig. \ref{fig:extend_NGSIM}(b).
\item Maintain a constant acceleration of $0.2 ~\mathrm{m/s^2}$ for vehicle $j$ until it reaches the new initial speed of the vehicle$i$, as illustrated in Fig. \ref{fig:extend_NGSIM}(b).
\end{enumerate}
This procedure ensures a continuous and realistic extension of the trajectories for simulation purposes. It is noted that the extended trajectory is primarily constructed based on real human driving data, with over 80\% of the trajectory segments originating from the NGSIM dataset. The extension is achieved by systematically connecting multiple human-driven trajectory fragments, which inherently reflect realistic driving behaviors across various traffic phases. This method allows us to create longer travel trajectories that preserve the temporal patterns observed in real-world traffic.

\section{Experimental Design and Configuration} \label{sec:design}
In this section, we provided an overview of the experiment design setup and methodologies used to evaluate the proposed framework within a simulated mixed-traffic environment. This section outlines the experiment design, experimental settings, and performance metrics employed to analyze the behavior and impact of automated vehicles alongside traditional human-driven vehicles.

\subsection{Experiment Design}
The experiments were conducted on the MATLAB platform, where 100 independent simulations of mixed traffic flow were executed using different random seeds to ensure statistical validity. Each simulation lasted 20 minutes and involved 100 vehicles, with randomly selected and sequenced lead-vehicle trajectories. The reported results represent the averaged values across all simulations, and the variations among repetitions were found to be small, indicating the robustness and consistency of the observed trends. Furthermore, the composition of the vehicle types, CAVs, CHVs, and THVs, was randomly assigned to the 100 vehicles in accordance with their respective MPRs. It is noted that the AVs refer specifically to highly automated vehicles capable of full self-driving functions (e.g., Level 4-5 automation), rather than vehicles equipped with Advanced Driver Assistance Systems (ADAS) like ACC. While ACC and other ADAS technologies are indeed widely adopted in recent vehicle models, these systems still rely heavily on human supervision and cannot be considered fully automated in the context of this research. 

\begin{figure}[!ht]
  \centering
  \includegraphics[width=0.5\textwidth]{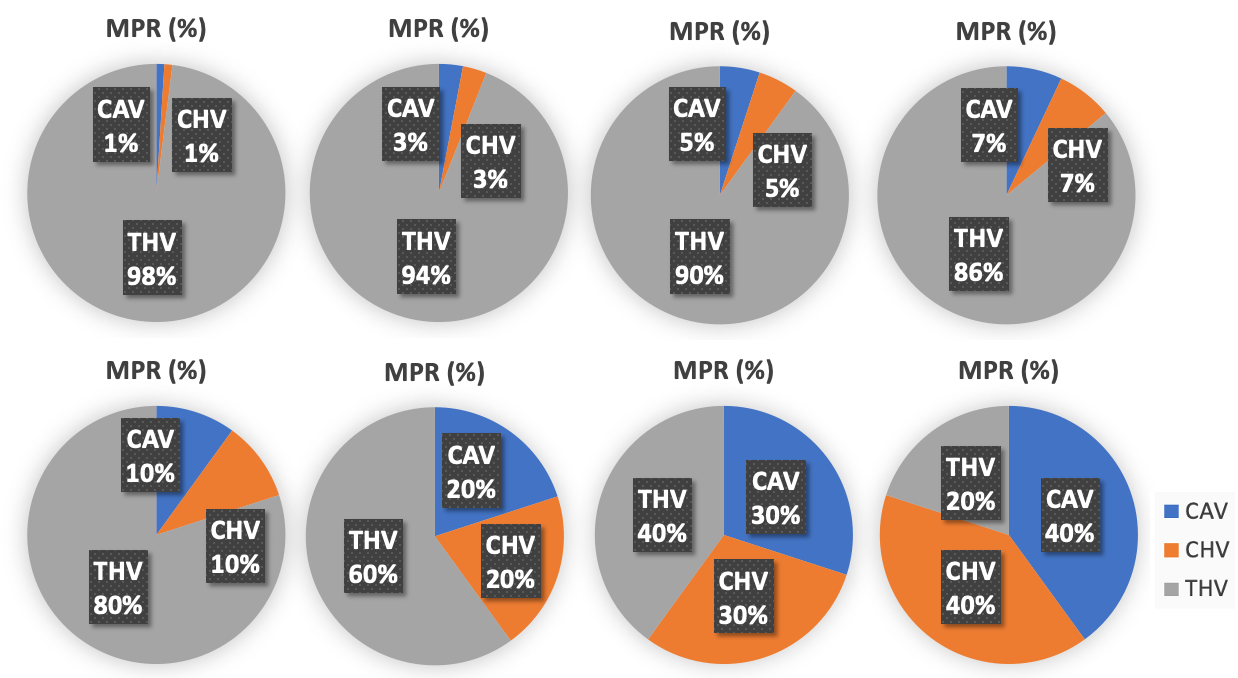}
  \caption{Traffic composition of different vehicles in different market penetrations }\label{fig:MPR}
\end{figure}

The experiments in this study are designed to evaluate the effects of cooperative platooning utilization in the early stages of CAV implementation, with a MPR of CAVs up to 40\%. We assume that connectivity is deployed at a higher rate than automation. Thus, the MPR of AVs is less than that of CVs (CAVs + CHVs). Specifically, as the percentage of CVs increased from 2\% to 80\%, the percentage of CAVs increased from 1\% to 40\%, with the percentage of CHVs matching that of the CAVs, as shown in Fig. \ref{fig:MPR}. For all figures in Section \ref{sec:results}, the x-axis labels denote the traffic composition. 

\begin{figure}[!ht]
  \centering
  \includegraphics[width=0.5\textwidth]{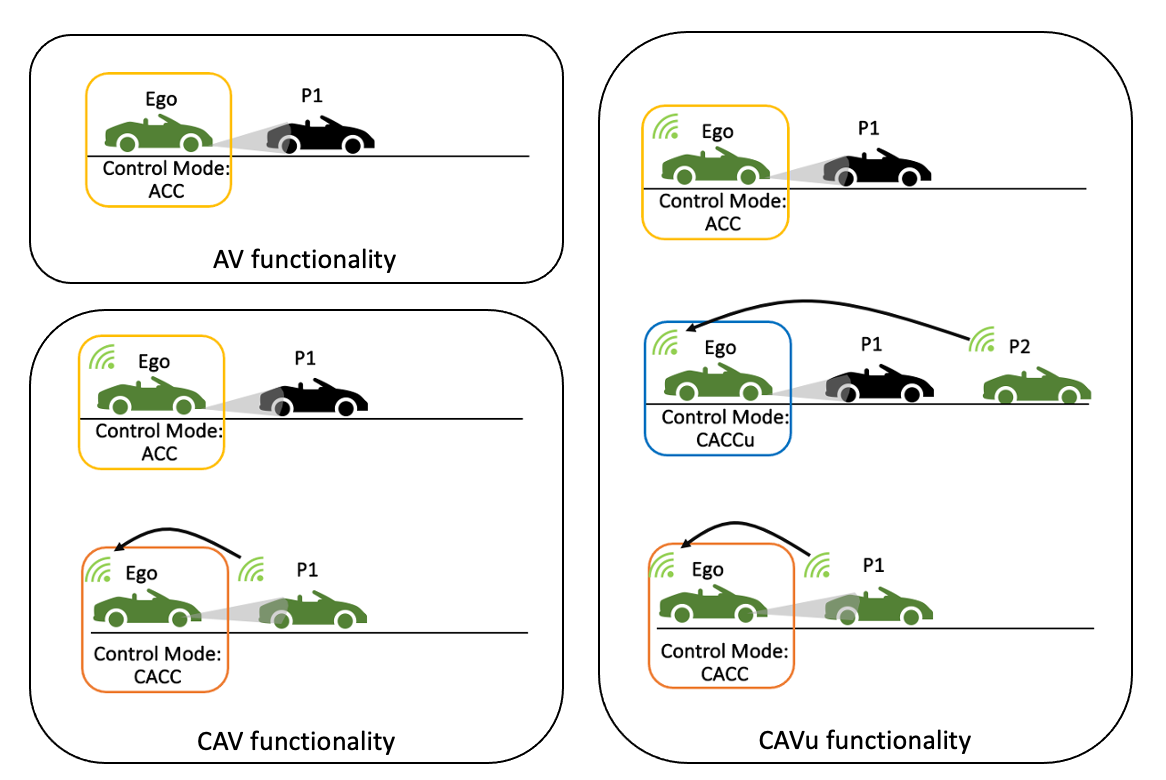}
  \caption{Automated Vehicle Functionality and Corresponding control planner}\label{fig:CAV}
\end{figure}

To evaluate the impacts of cooperative platooning in mixed traffic flow, four functionalities of automated vehicles involving three control modes are simulated, as illustrated in Fig. \ref{fig:CAV}. The details of each type are as follows:
\begin{itemize}
\item CAVu with Lane Change functionality (CAVu-LC): These CAVs are equipped with ACC, CACC, and CACCu systems. The SVIS can identify the connectivity of nearby vehicles, including the second vehicle ahead, and enable lane changes for cooperative platooning.
\item CAVu without Lane Change functionality (CAVu): These CAVs are equipped with ACC, CACC, and CACCu systems. The SVIS can identify the connectivity of nearby vehicles, including the second vehicle ahead; however, lane changes for cooperative platooning are not facilitated.
\item CAV: These CAVs are equipped with ACC and CACC systems. The SVIS can identify the connectivity of the surrounding vehicle.
\item AV: These are basic automated vehicles equipped only with ACC systems.
\end{itemize}

\subsection{Simulation Settings and Parameters}

For automated vehicles, the realistic distance and velocity errors of GPS and radars were generated based on a normal distribution \cite{sensor}, with a mean of zero and standard deviations detailed in Table \ref{tab:parm-CAV}. According to studies, the GPS velocity error is negligible \cite{GPS_error1, GPS_error2}. Additionally, the parameters for the control planner and the vehicle dynamics model of the CAV are outlined in Table \ref{tab:parm-CAV}. It is important to note that the control gains and desired time headways across different control planners are standardized to ensure a fair comparison.

\begin{table}[h]
	\caption{Parameters Settings for Connected Automated Vehicles}\label{tab:parm-CAV}
	\begin{center}
	\begin{threeparttable}
		\begin{tabular}{l r}
		    \hline
		    \hline
	        Parameter Name (Unit) & Value/Range \\
                \hline
	        \multicolumn{2}{c}{Radar and GPS Error}\\
	        Radar Distance Error $(m)$ & Normal $\mathcal{N}(0,0.1)$ \\
	        Radar Velocity Error $(m)$ & Normal $\mathcal{N}(0,0.1)$ \\
	        GPS Distance Error $(m)$ & Normal $\mathcal{N}(0,1)$ \\
	        Search Distance Region & Chi-squared $\chi^2(2,0.011)$ \\
	        Search Speed Region & Chi-squared $\chi^2(1,0.0049)$ \\
	        Maximum Error Rate ($Er_{max}$) & $10^{-8}$ \\
	        Connectivity Search Inner Step $(n)$ & 34 \\
	        Connectivity Search Step $(k)$ & 6 \\
         \hline
	        \multicolumn{2}{c}{control planner}\\
	        Proportional Gain ($k_p$) & 0.3 \\
	        Derivative Gain ($k_d$) & 0.7 \\
                Desired Time Headway ($T,s$) & 1.2 \\
                Human gains (CACCu) ($\alpha,\beta$) & 0.76, 0.51 \\
                Time Headway (CACCu) ($T_{ovm},s)$ & 0.57 \\
                \hline
                \multicolumn{2}{c}{Vehicle Dynamics Model }\\
                Power Train ($\tau$) & 0.5 \\
                System Delay ($t_d, s$) & 0.3 \\
	        \hline
	        \hline
		\end{tabular}
        \end{threeparttable}
	\end{center}
\end{table}

\begin{table}[h]
	\caption{Parameters Settings for Human Model}\label{tab:parm-IDM}
	\begin{center}
	\begin{threeparttable}
		\begin{tabular}{l r}
		    \hline
		    \hline
	        Parameter Name (Unit) & Value/Range \\
	        \hline
                Desired Speed ($v_d, m/s$) & 25 \\
                Minimum Gap ($s_0,m$) & 2 \\
                Maximum Acceleration ($a,m/s^2$) & 4 \\
                Comfort Deceleration ($b,m/s^2$) & 4 \\
                Emergency Deceleration ($b_e,m/s^2$) & 8 \\
                Safe Time Headway  ($T,s$) & Uniform $U\sim[1,2]$ \\
                Human Delay ($k, s$) & 0.9 \\
                TTC Threshold ($TTC, s$) & 3.6 \\
                Relative Distance Error ($v_s$) &  0.05\\
                Relative Approach Rate ($\sigma_r$) & 0.01\\
                Persistence ($\tau, s$) &  20\\
	        \hline
	        \hline
		\end{tabular}
        \end{threeparttable}
	\end{center}
\end{table}

For the human driver model, the parameters of the IDM are detailed in Table \ref{tab:parm-IDM}. To evaluate lane changes for cooperative platoons of CAVs, human drivers in this study do not perform lane changes. For safety concerns, the predefined TTC value is set at 3.6 seconds, based on experiments cited in \cite{Raymond2006Time}. The emergency deceleration is established at -8 $\mathrm{m/s^2}$, as recommended by \cite{Kudarauskas2007Analysis}. To accurately represent human drivers in the simulation, we tested various reaction delays using 100 pairs of vehicles from the NGSIM dataset. A reaction delay of 0.9 seconds was identified as the most suitable parameter to mimic real-world NGSIM traffic flow.

\subsection{Key Performance Metrics} \label{sec:performance_metrics}
The simulation results are analyzed and compared using average performance metrics for both CAVs and all vehicles within the traffic. To assess cooperative platooning utilization, safety, driving comfort, efficiency, and fuel efficiency of CAVs, the following key metrics are analyzed:
\begin{itemize}
\item Cooperative Platooning Utilization: This metric calculates the percentage of CACC and CACCu utilization. 
\item Maximum Unsafe Spacing Error: This metric represents the absolute maximum spacing error, occurring when a CAV is closer to the preceding vehicle than desired.
\item Acceleration RMS: The root mean square (RMS) of accelerations is calculated to evaluate vehicular stability.
\item Minimum Speed: This metric measures the minimum speed of the vehicles during the simulation.
\item Fuel: Average fuel consumption is calculated using a power-based microscopic fuel consumption model, as detailed in \cite{Virginia2011Hesham}.
\end{itemize}
For the evaluation of mixed traffic, driving comfort, fuel efficiency, and traffic efficiency (as indicated by minimum speed) within traffic flow are analyzed.

\section{Simulation Results and Analysis} \label{sec:results}
In this section, we evaluated the efficiency and accuracy of SVIS, the impacts of mixed cooperative platooning and lane changes for cooperative platoons across various MPRs of CAVs. The evaluation of the proposed control framework includes analyzing specific performance metrics, as detailed in Section \ref{sec:performance_metrics}.

\subsection{Vehicle Identification Evaluation}

Given that the accuracy and efficiency of the identification system influence the cooperative platoon rate and the associated benefits, we evaluated the accuracy and response time of the SVIS and its impacts on system-level performance. 

Simulation results showed that the SVIS required, on average, 4.5 seconds to identify surrounding connected vehicles and initiate CACC. As a result, in 20-minute simulations, CAVs that successfully established a connection transitioned from ACC to CACC and remained in cooperative mode for over 99\% of the simulation duration. Notably, the misidentification error, defined as the proportion of incorrectly identified connected vehicles, was zero across all traffic scenarios, indicating that no CAVs formed a platoon with an incorrect target.

\begin{figure}[t]
  \centering
  \includegraphics[width=0.5\textwidth]{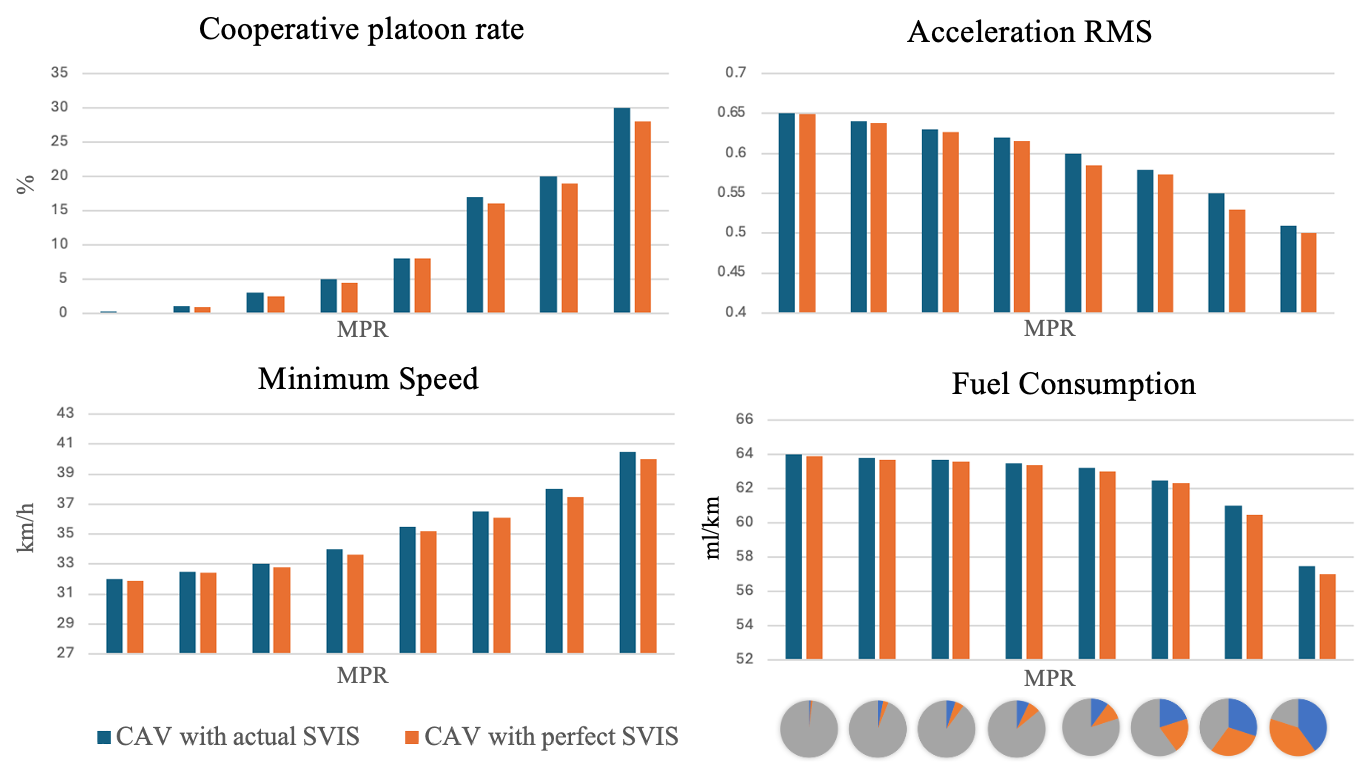}
  \caption{Performance comparison between CAVs using the SVIS and those with a perfect SVIS system capable of identifying connected vehicles instantly and with complete accuracy.}\label{fig:SVIS_sim}
\end{figure}

However, approximately 6\% of connected preceding vehicles were mistakenly classified as unconnected, preventing some CAVs from engaging in cooperative behavior because of GPS and sensor inaccuracies. To assess the impact of such misclassification on system-level performance, a perfect SVIS (which identifies vehicles instantly and with complete accuracy) is compared with the actual SVIS. As shown in Fig. \ref{fig:SVIS_sim}, the performance gap between the two systems peaked at a CV MPR of 60\%, with the largest differences observed in cooperative platoon formation rate, acceleration RMS, minimum speed, and fuel consumption. The difference indicates that the accuracy and efficiency of the identification system are critical to enabling effective cooperation. An accurate and responsive identification system is essential to minimize performance degradation caused by delays or misclassifications.

\subsection{Simulation Results: Mixed Cooperative Platooning} 

\begin{figure}[t]
  \centering
  \includegraphics[width=0.4\textwidth]{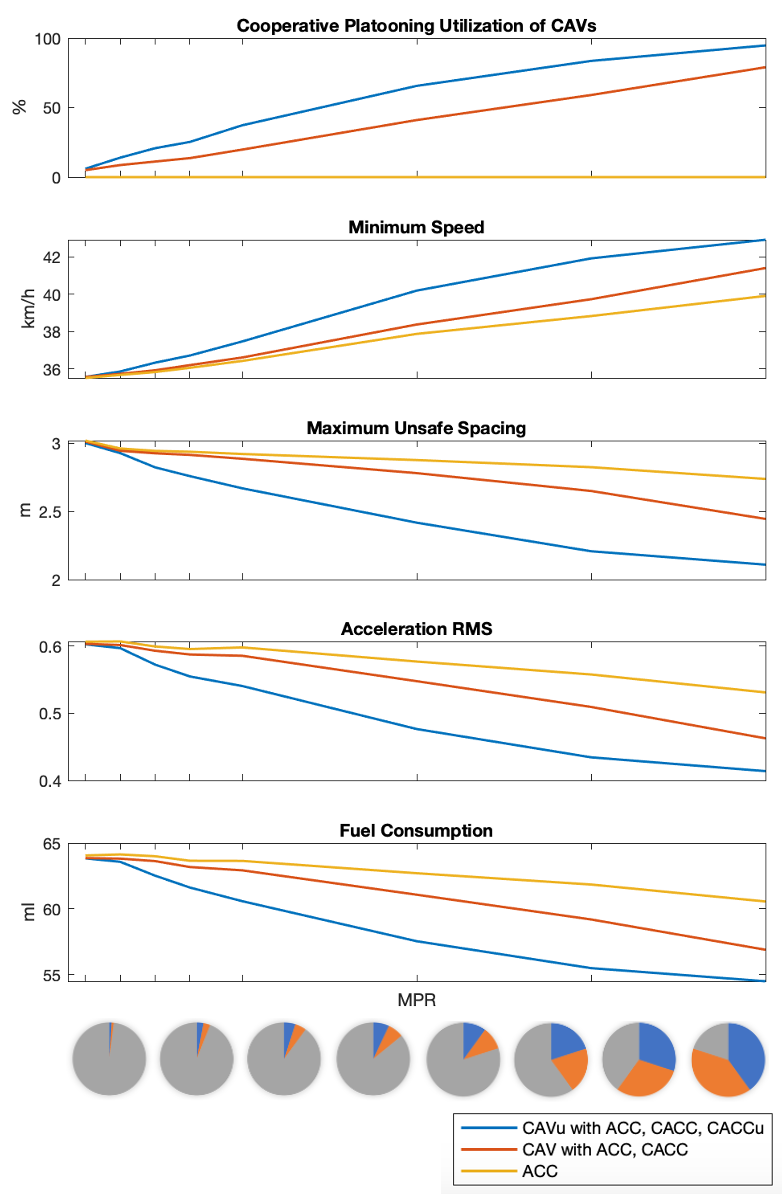}
  \caption{Performance of connected automated vehicles in different market penetration rates of connected vehicle}\label{fig:CAV_low}
\end{figure}

\begin{figure}[!ht]
  \centering
  \includegraphics[width=0.4\textwidth]{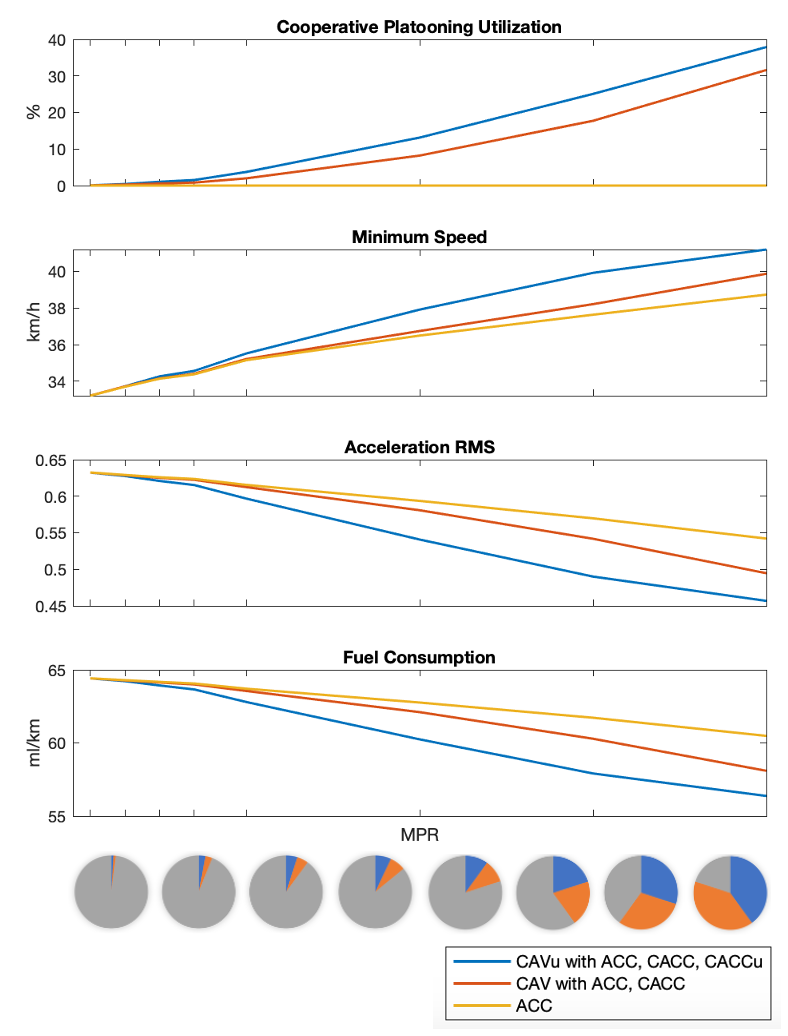}
  \caption{Performance of mixed traffic flow with different market penetration rate of connected vehicle}\label{fig:traffic_low}
\end{figure}

When CAVs are equipped with the proposed control framework, which includes ACC, CACC, and CACCu control planners but does not incorporate lane changes for cooperative platooning, they can form both cooperative and mixed cooperative platoons. We compared this proposed framework to two existing frameworks: one that supports cooperative platooning only with the immediate preceding vehicle and defaults to ACC when the preceding vehicle is unconnected, and another that exclusively uses the ACC system. Specifically, simulations were conducted to compare automated vehicles with three functionalities at low MPR: AVs functionality only with ACC, CAVs functionality with ACC and CACC, and CAVu functionality with ACC, CACC, and CACCu. As shown in Fig. \ref{fig:CAV}, each type of automated vehicle functionality was represented at rates ranging from 1\% to 40\%, with an equal percentage of CHVs, while the remainder of the vehicles were THVs. The simulation results, presented in Figs. \ref{fig:CAV_low} and \ref{fig:traffic_low}, illustrate the impact of mixed cooperative platooning on CAV performance and overall traffic conditions.

Compared to the AV functionality (baseline), which cannot utilize any cooperative platooning, the cooperative platooning utilization rate for CAV functionality increased from 0.0\% to 79.3\% as the MPR of CVs rose from 2\% to 80\%. As shown in Fig. \ref{fig:CAV_low}, significant improvements in risk reduction, comfort, and fuel consumption were observed when at least 20\% of the vehicles were CAVs and CV penetration reached 40\%. In scenarios with 80\% connected mixed traffic, the improvements achieved in risk mitigation, comfort, and fuel consumption were 16.9\%, 22.6\%, and 8.7\%, respectively.

When comparing the CAVu functionality with the conventional CAV and AV functionalities, as shown in Fig. \ref{fig:CAV_low}, notable differences emerge in cooperative platooning utilization. At a 10\% CV penetration rate, cooperative platooning utilization reached 9.5\% for CAV functionality, whereas CAVu functionality achieved 18.2\%, indicating a substantial initial enhancement over CAV functionality. This upward trend continued, with CAVu functionality exhibiting a steeper increase, resulting in 24\% greater utilization at a CV penetration rate of 60\% and a CAV rate of 30\%. Additionally, CAVu functionality demonstrated marked improvements over CAV functionality in key areas, including risk reduction, comfort, and fuel efficiency, with increases of 18.2\%, 36.8\%, and 6.9\%, respectively. These enhancements, attributed to mixed cooperative platooning, enable higher connectivity utilization. Notably, CAVu functionality begins delivering significant benefits in safety, comfort, and fuel efficiency at a CV MPR of 10\%, whereas CAV functionality requires a CV MPR of 40\% to yield comparable improvements.

In the overall traffic dynamics with CAV functionality, a similar trend in performance improvement is observed, as shown in Fig. \ref{fig:traffic_low}, though the improvement is relatively modest. This modest gain is attributed to the indirect positive effects on human drivers, who benefit from following the smoother speed trajectories of automated vehicles. Compared to the AV functionality baseline scenario, where automated vehicles are equipped solely with ACC systems, significant enhancements in comfort and fuel consumption are noted when the penetration of CAVs reaches 20\% and CV penetration is at 40\%. For mobility improvements, measured in terms of minimum required speed, a higher penetration of CAVs is necessary, with significant gains observed at a CAV penetration of 30\% and a CV penetration of 60\%. In scenarios involving CAVu functionality, significant improvements in comfort and fuel efficiency appear at a CV penetration rate of 20\%, while mobility improvements (measured by minimum speed) begin at a CV penetration rate of 40\%. These results indicated that CAVu functionality can achieve meaningful enhancements with 20\% fewer CVs than required by standard CAV functionality in mixed traffic scenarios.

\subsection{Simulation Results: Lane Changes} \label{sec:with_LC}
In the previous section, we demonstrated that mixed cooperative platooning can substantially improve CAV performance, especially in scenarios with low MPRs of CAVs. However, CACC generally offers greater advantages than CACCu, as CAVs employing CACCu must follow conventional human-driven vehicles, which are prone to human errors due to distractions and exhibit less predictable behavior. To mitigate this limitation, incorporating lane changes for CACC presents a promising solution to further optimize the benefits of cooperative platooning by enabling CAVs to align with connected vehicles. Therefore, in this section, we evaluate the impacts of the proposed control framework, which integrates a lane change model for cooperative platooning, within environments characterized by low CAV MPRs.

\begin{figure} [t]
    \centering
    \includegraphics[width=0.9\linewidth]{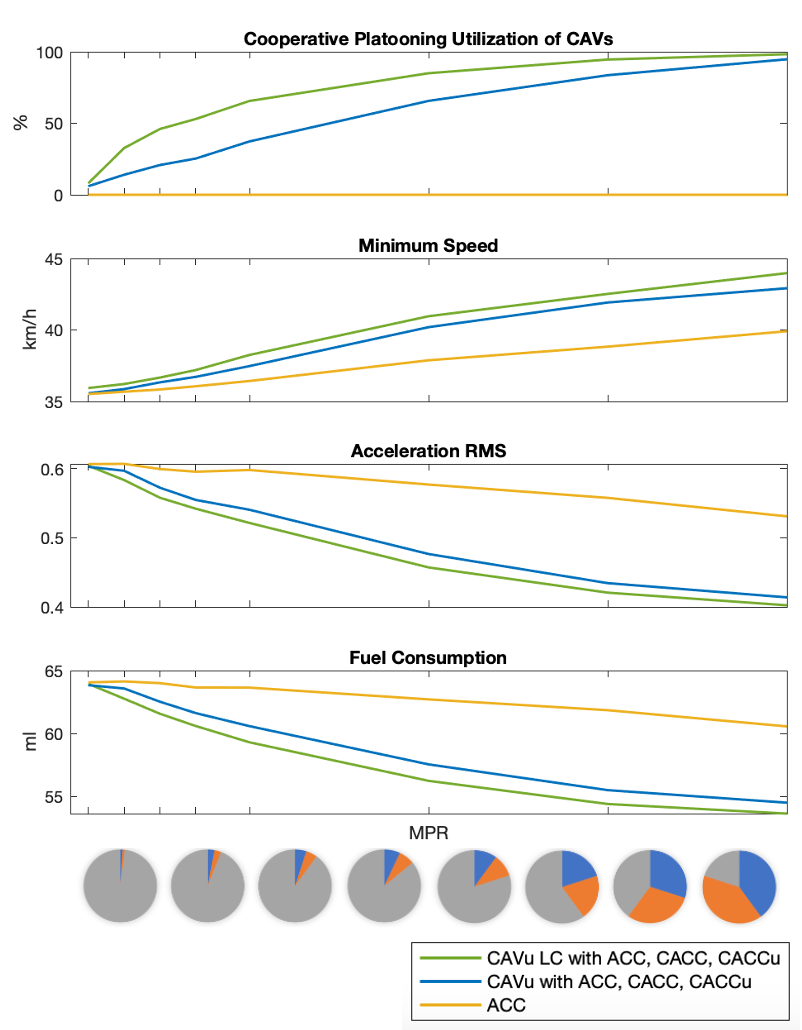}
    \caption{Automated vehicle performance comparison among AV, CACu and CAVu with strategic lane change functionality (CAVu-LC), under different market penetration rates of connected vehicles.}
    \label{fig:CAV_LC}
\end{figure}

\begin{figure} [ht!]
    \centering
    \includegraphics[width=0.8\linewidth]{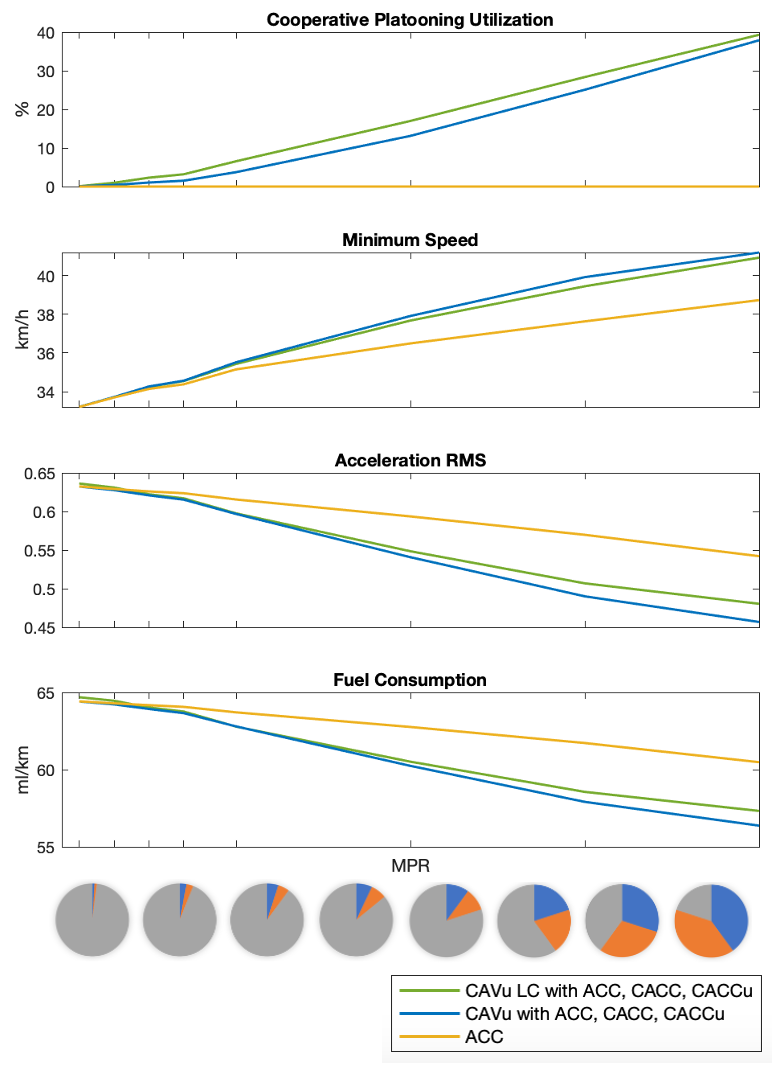}
    \caption{Mixed traffic flow performance comparison among AV, CAVu, and CACu with strategic lane change functionality (CAVu-LC), under different market penetration rates of connected vehicles.}
    \label{fig:traffic_LC}
\end{figure}

To assess the effects of lane-changing maneuvers on cooperative platooning within mixed traffic environments, we conducted an extensive series of experiments across various MPRs of CAVs. We compared the performance of CAVs and overall traffic flow employing lane-changing strategies to form cooperative platoons (CAVu-LC functionality) with that of various control frameworks: ACC functionality and the proposed CAVu functionality. Specifically, when the preceding vehicle is unconnected, the CAV explores opportunities to cooperate with nearby connected vehicles and decides whether to initiate a lane change to follow a connected vehicle in an adjacent lane. Fig. \ref{fig:CAV_LC} and  \ref{fig:traffic_LC} present the automated vehicle and traffic flow performance, respectively.

As shown in Fig. \ref{fig:CAV_LC}, the findings reveal that implementing lane changes for cooperative platooning significantly enhances the integration of CAVs into cooperative platoons, particularly at low market penetration rates (MPRs) of connected vehicles. Specifically, when lane changes are enabled for platooning, cooperative platoon utilization reaches approximately 35\% at a combined MPR of 6\% for CVs (3\% CHVs + 3\% CAVs). In contrast, scenarios without lane-changing capabilities (CAVu) achieve less than 10\% cooperative platoon utilization, highlighting the crucial role of strategic lane-changing maneuvers in optimizing cooperative system benefits at low connectivity levels.

Further analysis reveals that the CAVu-LC functionality provides additional performance improvements in terms of acceleration smoothness, speed consistency, and fuel consumption as the MPR of CAVs increases. Notably, these benefits show significant improvement beginning at a CV MPR of 6\% with the CAVu-LC functionality, whereas similar enhancements are observed only at a CV MPR of 10\% with the CAVu functionality. This indicates that the lane-changing capability in CAVu-LC allows for earlier and more effective integration into cooperative platoons, achieving smoother vehicle dynamics and enhanced fuel efficiency at lower connectivity thresholds.

Recognizing that multiple lane changes can cause disturbances to traffic flow and potentially diminish the benefits of forming cooperative platoons, we conducted a comprehensive analysis to investigate the broader implications of increased lane-changing activity within mixed traffic contexts. This analysis examines how lane changes impact the overall traffic system, balancing the advantages of cooperative platooning with the potential disruptions caused by excessive maneuvering. As shown in Fig. \ref{fig:traffic_LC}, this analysis specifically examines the effects of cooperative platoon formation through lane changes on traffic and energy efficiency. The results indicate that, at lower combined MPRs of CAVu-LC (below 60\%, with 30\% CAVs and 30\% CHVs), the impact of lane changes on overall traffic flow and energy efficiency is minimal. This suggests that enabling lane changes for cooperative platooning can significantly improve the operational efficiency of CAVs without causing major disruptions in mixed traffic environments. However, when the combined MPR of CVs exceeds 60\%, the increase in lane changes may lead to traffic disturbances, potentially offsetting some efficiency gains. These findings underscore the potential of lane-changing strategies to enhance CAV performance in cooperative platooning while maintaining a balanced and stable traffic flow within mixed traffic systems.

\section{Conclusions and Future Research} \label{sec:conclusion}

In the early stages of CAV deployment, limited opportunities exist for CAVs to engage in cooperative platooning. To address this, we propose a control framework designed to maximize cooperative platooning through mixed cooperative platooning and lane changes for cooperative platoons. This framework includes a vehicle identification system to detect connected leaders, a strategic lane change decision model, and multiple control planners, incorporating CACCu functionality in addition to CACC and ACC.

We evaluated the impacts of surrounding vehicle identification systems, mixed cooperative platooning, and lane changes in a microsimulation environment for mixed traffic, incorporating realistic factors such as GPS and sensor errors, human driving models with human factors, a vehicle dynamics model, and extended real-world data. Various aspects of CAV and traffic flow performance, including driving comfort, safety, fuel consumption, and traffic flow efficiency, were assessed across different MPRs of CAVs and CHVs within the traffic stream.

To initiate cooperative platooning, accurately identifying surrounding connected vehicles is the first critical step for establishing connectivity and enabling coordination. Simulation results indicate that the proposed vehicle identification system requires, on average, 4.5 seconds to identify nearby connected vehicles. While no target connected vehicles are mistakenly identified as incorrect connected vehicles, approximately 6\% of connected vehicles are incorrectly classified as unconnected, resulting in an unusability rate that undermines the effectiveness of cooperative platooning in mixed traffic. As the MPR of CVs increases, this degradation becomes more pronounced, with the adverse impact on platooning performance. These findings suggest that failing to account for vehicle identification accuracy could lead to an overestimation of cooperative platooning benefits in mixed traffic environments. An efficient vehicle identification system is therefore crucial for establishing reliable cooperation and maximizing the advantages of cooperative platooning.

In experiments conducted with low MPRs of CAVs up to 40\%, we investigated the impacts of mixed cooperative platooning (CACCu) and lane changes within cooperative platoons. The results showed significant improvements in CAV (CACC + ACC) functionality, particularly when the MPR of CVs exceeded 40\%, compared to the baseline ACC functionality. For CAVu functionality (ACC + CACC + CACCu), even a 10\% penetration of CVs substantially enhanced the utilization of cooperative platooning, yielding notable improvements.

Furthermore, CAVu functionality significantly improved overall traffic performance when the MPR of CVs exceeded 20\%, demonstrating superior results compared to standard CAV functionality. Additionally, CAVs equipped with strategic lane change functionality showed that lane changes can provide substantial benefits within mixed cooperative platoons (CACCu), starting from a CV MPR of 6\%, without causing significant traffic disturbances, even at a CV MPR of 60\%.

The successful implementation and demonstrated benefits of CACC and CACCu highlight the necessity of developing a reliable system for accurately identifying connected vehicles to form cooperative platoons. One particularly critical and challenging component is the identification of the second preceding vehicle, which remains a nontrivial task requiring further research to fully realize the advantages of CACCu. In this study, the lane change process is simplified, and cooperative interactions between CAVs during lane changes are not explicitly modeled. However, cooperative lane changes have the potential to enhance both safety and traffic stability by reducing disturbances and improving coordination. Therefore, future research should focus on developing and evaluating cooperative lane change strategies. Such investigations will significantly advance CAV capabilities, enabling more efficient, safer, and more coordinated traffic flow. Furthermore, while the proposed CAV framework is validated through simulation, field evaluation is equally essential and warrants further exploration to assess real-world applicability and effectiveness.

\section*{Acknowledgements}
This work is supported by the National Science Foundation under Grant No. (CMMI-2009342) and the Toyota Motor North America R\&D.

\bibliographystyle{IEEEtran}
\bibliography{interactapasample}


\begin{IEEEbiography}
[{\includegraphics[width=1in,height=1.25in,clip,keepaspectratio]{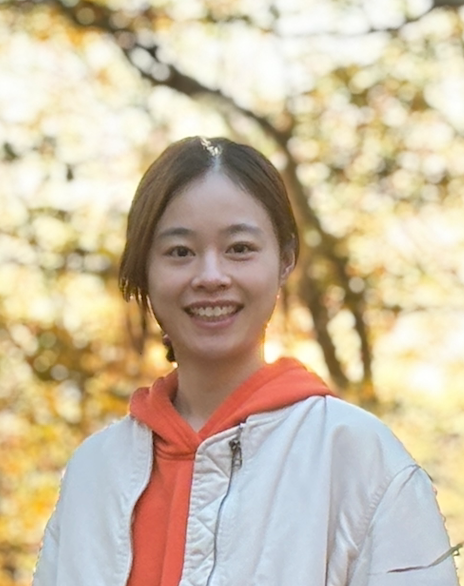}}]{Zeyu Mu} received her M.S. degree in Electrical Engineering from the University of Southern California in 2020. She is currently a Ph.D. candidate with the Link Lab and the Departments of Systems \& Information Engineering at the University of Virginia. Her research interests include control, optimization, and intelligent decision-making in connected and automated vehicles.
\end{IEEEbiography}

\vspace{-22pt}

\begin{IEEEbiography}
[{\includegraphics[width=1in,height=1.25in,clip,keepaspectratio]{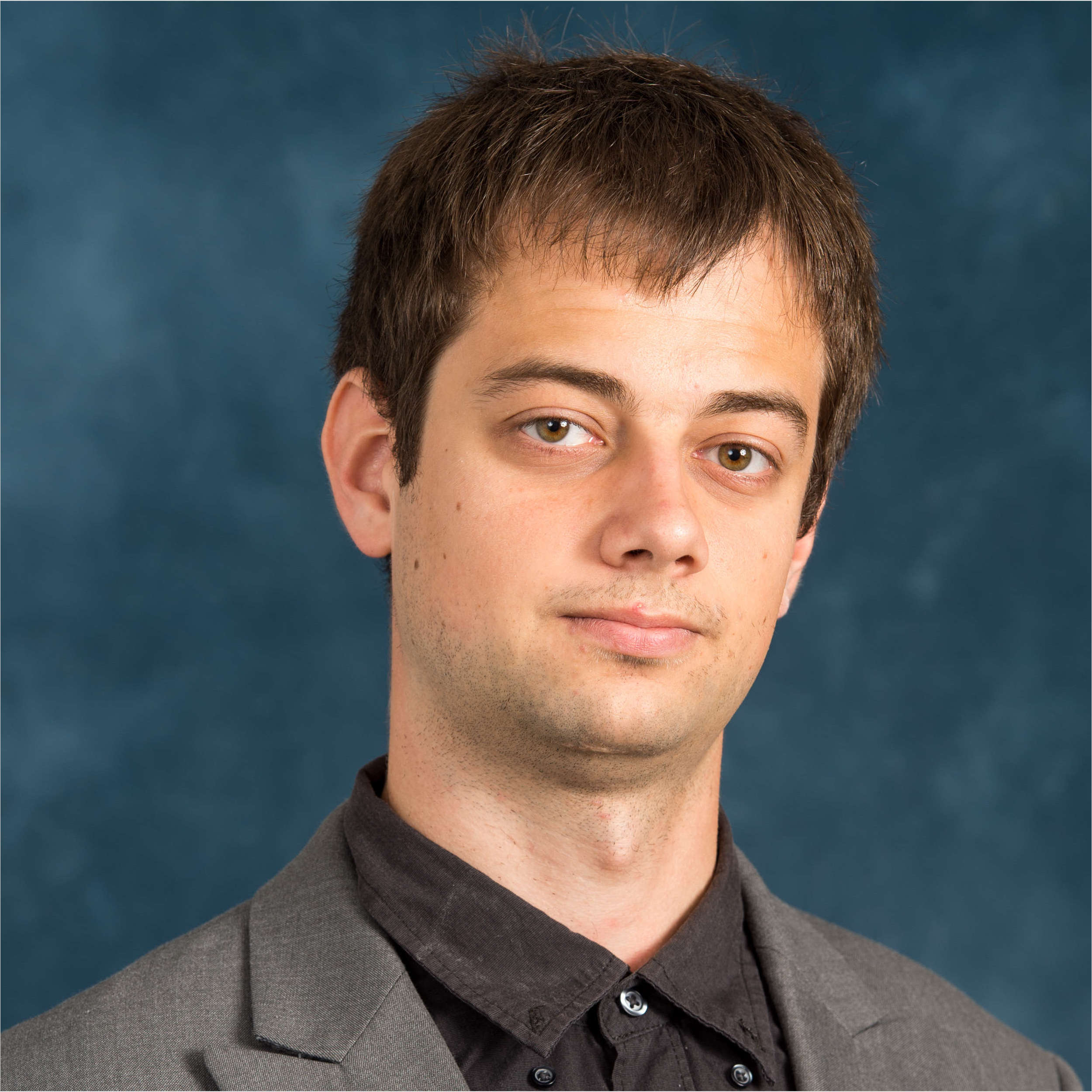}}]{Sergei Avedisov} works as a principal researcher at Toyota Infotech Labs. His research interests include cooperative automated driving, cooperative perception, cooperative maneuvering, platooning, and teleoperated driving. Sergei got his PhD in Mechanical Engineering from the University of Michigan in 2019.

\end{IEEEbiography}

\vspace{-33pt}

\begin{IEEEbiography}
[{\includegraphics[width=1in,height=1.25in,clip,keepaspectratio]{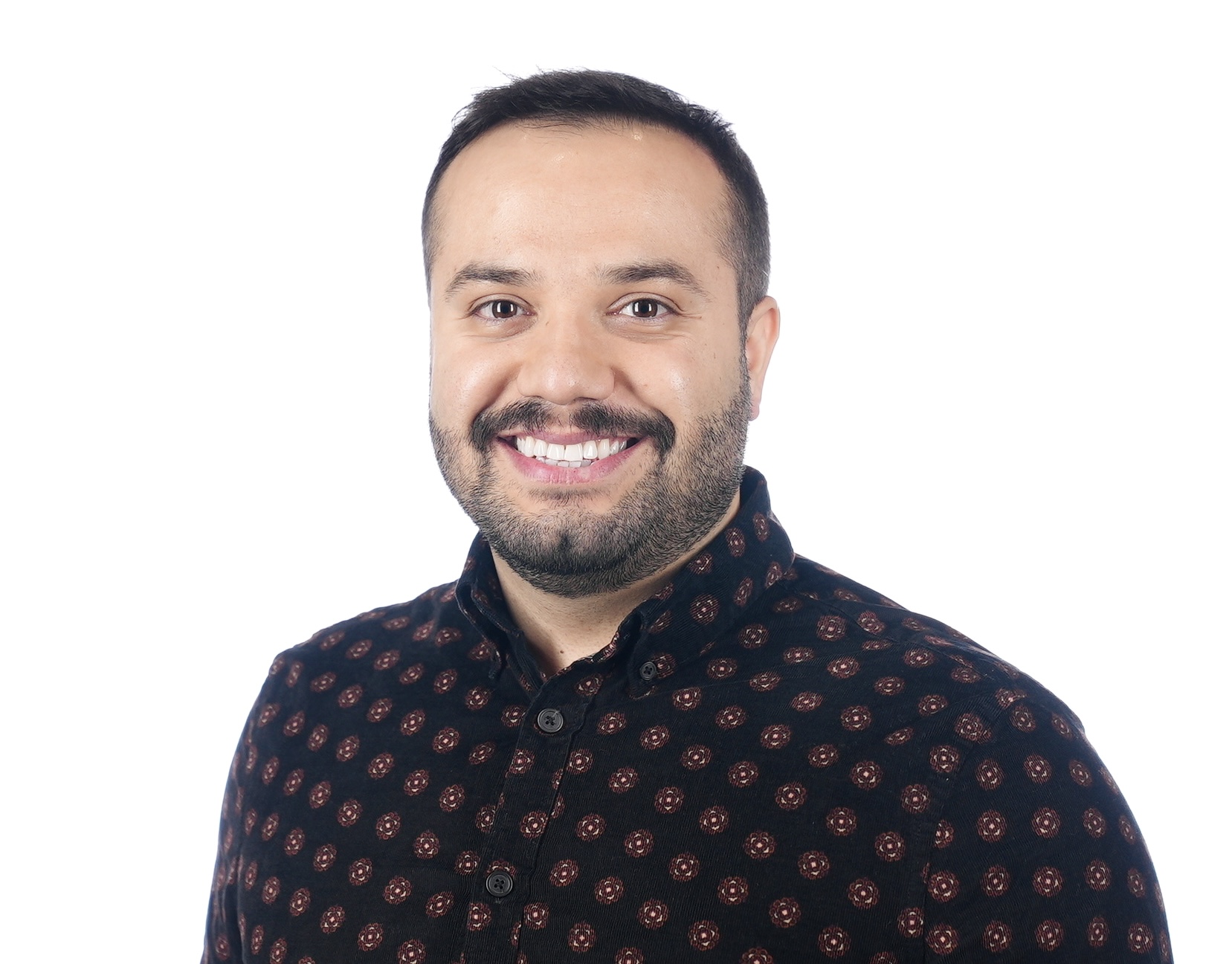}}]{Ahmadreza Moradipari} is a senior scientist at Toyota InfoTech Labs.  Dr. Moradipari received the B.Sc. degree in Electrical Engineering from Sharif University of Technology in 2017 and the M.Sc. and Ph.D. degrees from the University of California, Santa Barbara in 2019 and 2022, both in Electrical and Computer Engineering. His research interests are focused on designing algorithms for agents who learn from and control their environments.

\end{IEEEbiography}

\vspace{-33pt}

\begin{IEEEbiography}[{\includegraphics[width=1in,height=1.25in,clip,keepaspectratio]{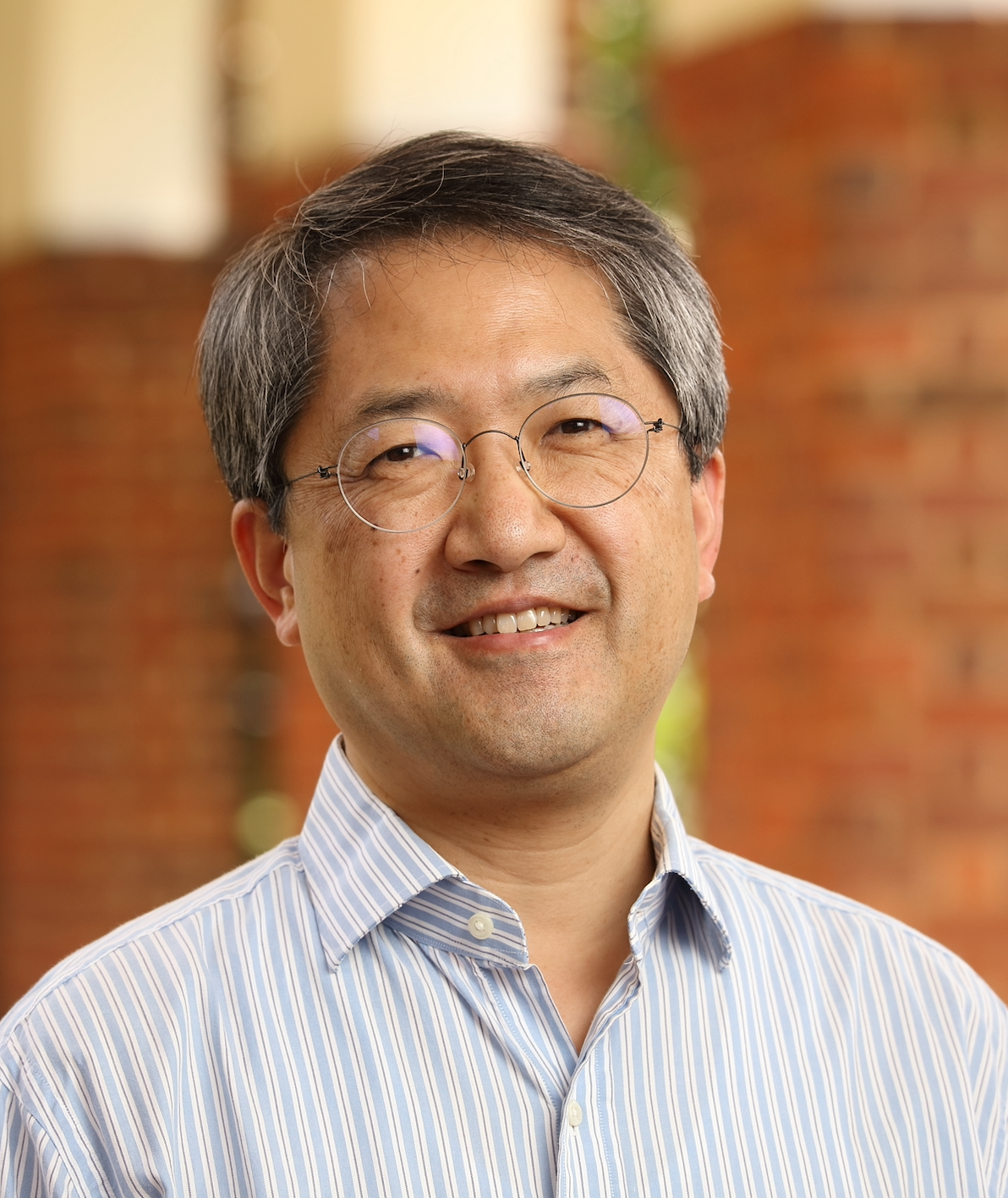}}]{Byungkyu Brian Park} (Senior Member, IEEE) is a Professor with the Link Lab and the Departments of Civil \& Environmental Engineering and Systems \& Information Engineering at the University of Virginia. He has published over 180 journal articles and conference papers on transportation system operations, management, and intelligent transportation systems. His research interests include cyber-physical systems for transportation, stochastic optimization, and connected and automated vehicles.
\end{IEEEbiography}

 




\vfill

\end{document}